\documentclass[aps, superscriptaddress, reprint, twocolumn]{revtex4-2}
\usepackage[T1]{fontenc}
\usepackage[utf8]{inputenc}

\usepackage{amsmath, amssymb, amstext, amsfonts}
\usepackage{mathtools}
\usepackage{dsfont}
\usepackage{graphicx}
\usepackage[dvipsnames]{xcolor}
\usepackage{float}
\usepackage{ragged2e}
\usepackage[caption=false, justification=justified]{subfig}
\DeclareCaptionJustification{justified}{\justifying}
\usepackage{placeins}
\usepackage{standalone}
\usepackage{amsthm}
\usepackage[colorlinks=true, linkcolor = BlueViolet, citecolor = BlueViolet, urlcolor = BlueViolet]{hyperref}
\usepackage{enumitem}

\usepackage{microtype}
\usepackage{times}

\usepackage{subdepth} 

\theoremstyle{definition}

\newtheorem{theorem}{Theorem}

\newtheorem{proposition}[theorem]{Proposition}

\newtheorem{statement}[theorem]{Statement}

\DeclareMathOperator{\Tr}{Tr}
\DeclareMathOperator{\sinc}{sinc}
\newcommand{\proj}{\hat{\Pi}}

\newcommand{\idop}{\hat{\mathds{1}}}
\newcommand{\diff}{\mathrm{d}}

\newcommand{\dos}{\mathcal{N}}
\newcommand{\tildos}{\widetilde{\mathcal{N}}}
\newcommand{\kraus}[2]{\hat{\mathcal{K}}_{#1}(#2)}
\newcommand{\krausd}[2]{\hat{\mathcal{K}}_{#1}^\dagger(#2)}

\newcommand{\maxds}{D_{\max}}

\newcommand{\func}{\dos}
\newcommand{\tilfunc}{\tildos}
\newcommand{\cue}{\text{CUE}}
\newcommand{\funcset}{\mathcal{F}}
\newcommand{\funcsubset}{\mathcal{F}_p}
\newcommand{\scrtimef}{\lambda}
\newcommand{\invscrtimef}{\lambda^{-1}}
\newcommand{\subsystem}[1]{\mathcal{H}_{#1}}

\newcommand{\speedL}{L}

\newcommand{\maintextEqCPdef}{2}

\newcommand{\edit}[1]{{#1}} 
\newcommand{\editb}[1]{{#1}}

\begin{document}

\title{Exact universal bounds on quantum dynamics and fast scrambling}
\author{Amit Vikram}
\affiliation{Joint Quantum Institute and Department of Physics, University of Maryland, College Park, MD 20742, USA}
\author{Victor Galitski}
\affiliation{Joint Quantum Institute and Department of Physics, University of Maryland, College Park, MD 20742, USA}

\begin{abstract}
Quantum speed limits such as the Mandelstam-Tamm or Margolus-Levitin bounds offer a quantitative formulation of the energy-time uncertainty principle that constrains dynamics over short times. We show that the spectral form factor, a central quantity in quantum chaos, sets a universal state-independent bound on the quantum dynamics of a complete set of initial states over arbitrarily long times, which is tighter than the corresponding state-independent bounds set by known speed limits. This bound further generalizes naturally to the real-time dynamics of time-dependent or dissipative systems where no energy spectrum exists. We use this result to constrain the scrambling of information in interacting many-body systems. For Hamiltonian systems, we show that the fundamental question of the fastest possible scrambling time -- without any restrictions on the structure of interactions -- maps to a purely mathematical property of the density of states involving the non-negativity of Fourier transforms. We illustrate these bounds in the Sachdev-Ye-Kitaev model, where we show that despite its ``maximally chaotic'' nature, the sustained scrambling of sufficiently large fermion subsystems via entanglement generation requires an exponentially long time in the subsystem size.
\end{abstract}

\maketitle

\textit{Introduction}--- The energy-time uncertainty principle sets fundamental limits on the speed of quantum dynamical processes. Specific formulations of this principle~\cite{MT, AAMT, ML, LevitinToffoli,QSLreview1, flippedML, GongHamazakiNoneqbBounds}, such as the Mandelstam-Tamm (MT)~\cite{MT, AAMT} and Margolus-Levitin (ML)~\cite{ML, flippedML} bounds (see also Refs.~\cite{QSLreview1, GongHamazakiNoneqbBounds} for reviews and extensions), are expressed in terms of a \textit{single} parameter $\Delta_E$ characterizing e.g. the standard deviation of energy (MT) or the mean energy relative to the ground state (ML), in a given initial state. In general, these allow a given decay in the return probability of the state only after a sharp time (proportional to $\Delta_E^{-1}$). However, such sharp bounds do not provide useful constraints on many-body processes that typically occur over timescales much larger than $\Delta_E^{-1}$, such as the thermalization of interacting many-body systems.

Thermalization has been at the focus of several developments in nonequilibrium statistical mechanics~\cite{PolkovnikovNoneqbReview, Reimann2016, EisertShortReview, GongHamazakiNoneqbBounds}, many-body quantum chaos~\cite{deutsch1991eth, srednicki1994eth, rigol2008eth, DAlessio2016, Borgonovi2016, deutsch2018eth, ChanScrambling, ProsenErgodic, ClaeysErgodicCircuits, ArulCircuit, ProsenThermalization} and quantum information~\cite{ChanScrambling, ProsenErgodic, ClaeysErgodicCircuits, ArulCircuit, ProsenThermalization, GoogleScrambling}. In our current understanding, many-body thermalization is driven by the generation of a high degree of entanglement between the interacting particles~\cite{tumulka_CT, CanonicalTypicalityPSW, NormalTypicality, subETH, AbaninMBL}.
The question of how fast this entanglement can be generated, irrespective of any restrictions on the nature of interactions, has come to be of fundamental interest, partly motivated by a conjectured correspondence between the black hole information problem and a form of thermalization called the \textit{scrambling} of information in Hamiltonian many-body systems~\cite{HaydenPreskill, SekinoSusskind, LashkariFastScrambling, BentsenGuLucasScrambling, LucasEntanglementVsOTOC}. On the other hand, useful many-body speed limits known so far require highly restrictive assumptions such as spatially local interactions~\cite{LiebRobinson, NachtergaeleLRbound, HastingsLRbound, GongHamazakiNoneqbBounds} or limited external control parameters~\cite{BukovSelsPolkovnikov}, preventing an exact solution of this problem in a general setting.

In this work, we derive a universal bound on quantum dynamics by considering the evolution of a complete set of coarse-grained initial states, e.g., a complete set of states for a subsystem of particles. This bound is directly given in terms of the spectral form factor~\cite{Haake} (SFF; see Eq.~\eqref{eq:SFFdef}) that characterizes the full profile of the energy spectrum in Hamiltonian systems, and also generalizes to non-Hamiltonian systems. We use this bound to constrain the speed of scrambling of information within subsystems of a many-body system. For Hamiltonian systems, we argue that any subsystem can typically \textit{remain} scrambled for a \edit{sustained} length of time only after an asymptotically long scrambling time in the subsystem size. Our bound constrains the ``\edit{sustained}'' scrambling time of a particular Hamiltonian system in terms of its density of states. Finally, we map the problem of bounding the fastest scrambling time to a purely mathematical statement, which in turn is related to the as yet unresolved mathematical problem of the necessary asymptotic conditions for a sufficiently well-behaved function (related to the SFF) to have a non-negative Fourier transform (related to a physical constraint on the density of states to be non-negative).

\begin{figure}[!b]
\includegraphics[width=\columnwidth]{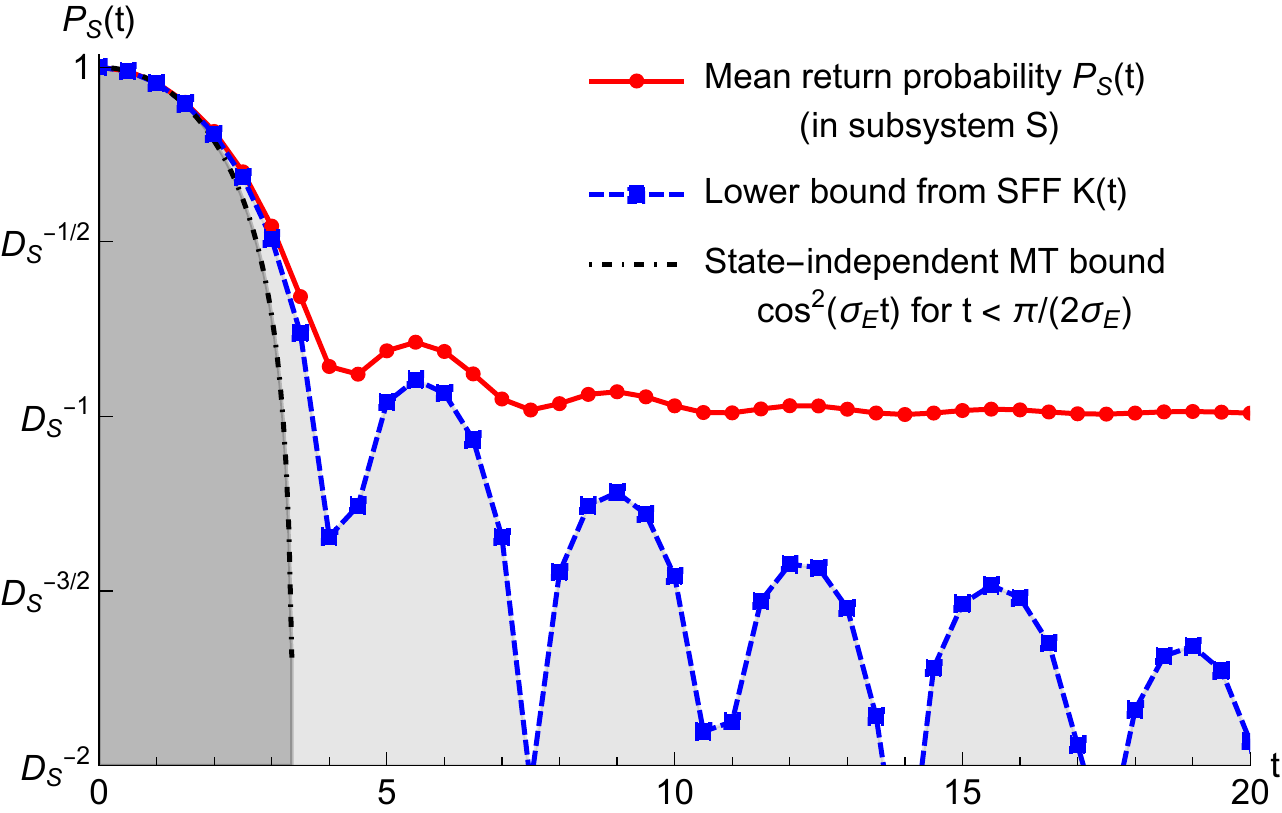}
\caption{\edit{Illustration (log-linear) of scrambling speed limits via Eqs.~\eqref{eq:dynamicalinequality}, \eqref{eq:sffcosbound}, \eqref{eq:ps_scrambling} and \eqref{eq:kt_scrambling} in a realization of the SYK-$4$ model of randomly interacting fermions, for initial fermion Fock states given by Eq.~\eqref{eq:SYKinitialstates} in a subsystem of $N_S = 7$ sites ($D_S = 128$) among $N=10$ total sites ($D=1024$). The power-law decay of $K(t)$ (whose nontrivial effect is evident here up to the first ``peak'' at $t\approx 6$) leads to exponentially slow sustained scrambling via Eq.~\eqref{eq:ts_SYK} as $N, N_S \to \infty$.}}
\label{fig:SYK}
\end{figure}

Correspondingly, our main results are: 1.~Eq.~\eqref{eq:dynamicalinequality}, providing a universal bound on quantum dynamics with relevance to long times, 2.~Eq.~\eqref{eq:kt_scrambling}, which supplies a rigorous necessary condition on the SFF to allow scrambling at a given time, and 3. Eq.~\eqref{eq:ts_dos} that relates the fastest allowed \edit{sustained} scrambling time of Hamiltonian many-body systems to the mathematical properties of \edit{its density of states, leading to Statements \ref{statement:gaussdos} and \ref{statement:sharpdos} describing universal constraints on fast scrambling. These results are illustrated analytically (Eq.~\eqref{eq:ts_SYK}) and numerically (Fig.~\ref{fig:SYK}) for the Sachdev-Ye-Kitaev (SYK) model}. We  use the formal asymptotic notation~\cite{knuth1976asymptotic, NielsenChuang} in this Letter, with the symbols $\omega$, $\Omega$, $\Theta$, $O$, $o$ respectively representing the order-of-magnitude versions of $>, \geq, =, \leq, <$.

\textit{Setup, and quantities of interest}--- We consider a general quantum mechanical system with a Hilbert space $\mathcal{H}$ of dimension $D$ \edit{(e.g, $D = 2^N$ for $N$ qubits)}, whose state at time $t$ is specified by a density operator $\hat{\rho}(t)$. Given an initial state $\hat{\rho}_0$, its time evolution may be generated by a Hamiltonian $\hat{H}$,
\begin{equation}
    \hat{\rho}(t) = e^{-i\hat{H}t} \hat{\rho}_0 e^{i\hat{H}t},
\end{equation}
or more generally, any linear dynamics of the form:
\begin{equation}
    \hat{\rho}(t) = \hat{\mathcal{T}}^t [\hat{\rho}_0] \equiv \sum_{r=1}^{M} \kraus{r}{t} \hat{\rho}_0\krausd{r}{t}.
    \label{eq:timeevolution}
\end{equation}
Eq.~\eqref{eq:timeevolution} represents the most general completely positive linear quantum operation~\cite{NielsenChuang} with time-dependent Kraus operators $\kraus{r}{t}$, and accounts for both unitary and nonunitary e.g. dissipative dynamics. A restriction to $M = 1$ and $\kraus{1}{t} = \hat{U}(t)$ corresponds to unitary evolution generated by $\hat{U}(t)$, reducing further to Hamiltonian dynamics when $\hat{U}(t) = e^{-i\hat{H}t}$. We note that the $M=1$ case also allows for nonunitary ``regularized'' or ``filtered'' Hamiltonian evolution of the form $\kraus{1}{t} = g(\hat{H})e^{-i\hat{H}t}$, which has been of interest in some applications~\cite{ShenkerThouless, Reimann2016}.

A state- and observable-independent characteristic of such a generic quantum dynamical system is given by the generalized SFF:
\begin{equation}
    K(t) \equiv \frac{1}{D^2}\sum_{r=1}^{M}\left\lvert \Tr\left[\kraus{r}{t}\right]\right\rvert^2.
    \label{eq:SFFdef}
\end{equation}
For Hamiltonian dynamics, $K(t)$ is the Fourier transform of the $2$-point energy level correlation function satisfying
$0 \leq K(t) \leq K(0) = 1$, and was introduced in the study of quantum chaos~\cite{Haake}. Its late-time quantum fluctuations (usually of $O(D^{-1})$ magnitude)  play a central role in characterizing energy level statistics~\cite{BlackHoleRandomMatrix, ShenkerThouless, KosProsen2018, BertiniProsen, ChanExtended, SSS, ExpRamp1, ExpRamp2, ExpRamp3, ExpRamp4, MBLSFF, Lozej2021} and quantum dynamical ergodicity~\cite{dynamicalqergodicity}. Time-dependent and non-unitary analogues have additionally been considered in various contexts~\cite{timeDependentSFF, dissipativeFormFactor0, dissipativeFormFactor1, dissipativeFormFactor2, dissipativeFormFactor3}, and Eq.~\eqref{eq:SFFdef} also accounts for these cases. A key feature of the SFF is that in spite of its state- and observable-independence, it lends itself to direct experimental measurement in many-body systems through recently developed measurement protocols~\cite{SFFmeas, pSFF}. Our primary interest in this Letter is to obtain universal constraints on the dynamics of quantum thermalization in terms of this quantity.

To study  thermalization, we should consider states and physical observables. In our initial abstract setting, this role will be fulfilled by a complete set $S = \lbrace \proj_k\rbrace_{k=1}^{D_S}$ of $D_S$ orthonormal projection operators $\proj_k$, satisfying Hermiticity $\proj_k^\dagger = \proj_k$, orthonormality $\proj_k \proj_{\ell} = \delta_{k\ell} \proj_k$ and completeness $\sum_{k=1}^{D_S} \proj_k = \idop$. These may be interpreted as $D_S$ different coarse-grained initial states $\hat{\rho}_{0,k} = \proj_k /\Tr[\proj_k]$ of the system within $\mathcal{H}$, as well as observables $\proj_k$ measuring the probability of overlap of a given state with these coarse-grained states \edit{(called a projection valued measure~\cite{NielsenChuang})}. Such projectors have a notable merit for our present purposes: the eigenvalues of $\proj_k$ are uniquely fixed to be $1$ (with $\Tr\proj_k$-fold degeneracy) and $0$ (with $(D-\Tr\proj_k)$-fold degeneracy). This avoids complications arising from a non-universal set of eigenvalues of more conventional observables such as particle positions or spin/qubit states -- which serve  to \textit{label} specific measurement outcomes if $\hat{\mathcal{T}}^t$ is already specified -- and allows one to focus on the intrinsic aspects of quantum dynamics $\hat{\mathcal{T}}^t$ itself.

We focus on the \textit{mean return probability} $P_S(t)$ of the projectors at time $t$, noting that return probabilities are the main quantities of interest in the MT and ML bounds~\cite{QSLreview1, MT, AAMT, ML, LevitinToffoli}. This is given by:
\begin{equation}
    P_S(t) \equiv \frac{1}{D}\sum_{k=1}^{D_S} \Tr\left[\proj_k(t)\proj_k\right],
    \label{eq:psdef}
\end{equation}
where \edit{$\proj_k(t) = \Tr[\proj_k] \hat{\rho}_{0,k}(t)$ is the time-evolving initial state weighted by its degeneracy}. As shown later, $P_S(t)$ partly measures the failure of an initial state to thermalize (see also Eq.~\eqref{eq:ps_scrambling}), but the above is sufficient setup to state our first key result.


\textit{A universal dynamical inequality}--- In this general setting of a quantum system whose dynamics is given by Eq.~\eqref{eq:timeevolution} with the SFF defined by Eq.~\eqref{eq:SFFdef}, our primary result is the following inequality on the dynamics of the mean return probability defined in Eq.~\eqref{eq:psdef} for any set of projectors $S$ \edit{(see Fig.~\ref{fig:SYK})}:
\begin{equation}
    P_S(t) \geq K(t).
    \label{eq:dynamicalinequality}
\end{equation}
This is derived in the supplement~\cite{supplement} as a consequence of two simple inequalities: the contribution to $P_S(t)$ from each $\kraus{r}{t}$ cannot be less than the \textit{squared} mean return amplitude of any set of orthonormal pure states constituting the projectors under the action of $\kraus{r}{t}$, while this mean amplitude is further bounded from below by the magnitude of $\Tr[\kraus{r}{t}]$ due to the triangle inequality. The latter bound has been previously used in Ref.~\cite{dynamicalqergodicity}, in the restricted context of Hamiltonian (and Floquet) unitary dynamics, to characterize the ``aperiodicity'' of quantum dynamical systems in terms of their SFF; in that case, the bound is saturated in any basis of discrete Fourier transformed~\cite{dynamicalqergodicity} \edit{(more generally, complex Hadamard transformed~\cite{supplement}) energy} eigenstates. While other relations between suitable return probabilities and $K(t)$ can be derived by averaging over all operators in the entire Hilbert space~\cite{Reimann2016, ChaosComplexityRMT, CotlerHunterJones2, SSS} and losing basis-specific information (see Eq.~\eqref{eq:HaarScrambling}), or in the late-time regime of quantum fluctuations by typicality arguments or an ensemble average over $\hat{\mathcal{T}}^t$ for certain ``physical'' operators~\cite{argaman, ChalkerReturnProbability, WinerHydro}, Eq.~\eqref{eq:dynamicalinequality} is an \textit{exact} relation that holds for all time in \textit{any} given basis.

Eq.~\eqref{eq:dynamicalinequality} provides an \edit{alternative formulation} of the energy-time uncertainty principle for the dynamics of the most general completely positive quantum operation, \edit{introducing} a nontrivial sensitivity --- even in the Hamiltonian case --- to long times and microscopic values of $P_S(t)$ through the asymptotics of $K(t)$.
Its relation to the more familiar MT and ML bounds~\cite{MT, AAMT, ML, LevitinToffoli, flippedML, QSLreview1} in a Hamiltonian system is obtained through corresponding bounds on the SFF~\cite{SonnerThermodouble}. For instance, the MT bound states that the return probability of a single initial state with energy variance $\Delta_E^2$ cannot decay faster than $\cos^2 (\Delta_{E}t)$, which is related to Eq.~\eqref{eq:dynamicalinequality} via the following inequality~\cite{SonnerThermodouble}:
\begin{equation}
    K(t) \geq \cos^2(\sigma_E t),\ \text{\edit{for }}\lvert t\rvert < \frac{\pi}{2\sigma_E},
    \label{eq:sffcosbound}
\end{equation}
where $\sigma_E^2 = -K''(0)/2$ is the variance of the energy spectrum \edit{itself, rather than any particular state (see Fig.~\ref{fig:SYK})}. \edit{This is generalized in the supplement~\cite{supplement}, from which it follows that the bound of Eq.~\eqref{eq:dynamicalinequality} is tighter than \textit{state-independent} bounds on $P_S(t)$ stemming from combining known speed limits with the properties of the energy spectrum itself (or Kraus operators), such as $\sigma_E$ for a state-independent MT bound $P_S[\lvert t\rvert < \pi/(2 \sigma_E)] \geq \cos^2(\sigma_E t)$.
These form a class of bounds applying universally to \textit{any} mean return probability $P_S(t)$ in the system, among which Eq.~\eqref{eq:dynamicalinequality} is consequently the tightest}.
It is worth emphasizing a trade-off: while the MT and ML bounds apply to the return probability of \textit{individual} states instead of an average as in $P_S(t)$, working with $P_S(t)$ in Eq.~\eqref{eq:dynamicalinequality} has the advantage of capturing the full spectral and dynamical information encoded in $K(t)$ over all timescales. This advantage is crucial for the application we consider next.

\textit{Application to scrambling} --- Now, we specialize to many-body systems,  with an implicit $D\to\infty$ thermodynamic limit. Here, we consider subsystems consisting of a subset of the degrees of freedom e.g. spins/qubits, which corresponds to a factorization of $\mathcal{H} = \mathcal{H}_S \otimes \mathcal{H}_E$ into the $D_S$-dimensional subsystem Hilbert space $\mathcal{H}_S$ (e.g. $D_S = 2^{N_S}$ for a subsystem of $N_S$ qubits), and the effective $D_E = D/D_S$ dimensional component $\mathcal{H}_E$ of the system ``external'' to $\subsystem{S}$ formed by the remaining degrees of freedom, which can act as a thermalizing bath~\cite{tumulka_CT, CanonicalTypicalityPSW, NormalTypicality, subETH}. In this context, each projector in $S$ can be chosen to reduce to a given pure state in an orthonormal basis $\mathcal{B}_S = \lbrace \lvert k\rangle_S\rbrace_{k=0}^{D_S-1}$ for the subsystem, with $D_E^{-1}\Tr_E[\proj_k] = \lvert k\rangle_S\langle k\rvert$ (e.g., computational basis states of qubits within $\subsystem{S}$).

Following  Ref.~\cite{LashkariFastScrambling}, \editb{our criterion for} the subsystem $\subsystem{S}$ to be scrambled at time $t$, in a given basis $\mathcal{B}_S$, \editb{requires} \textit{all} basis states \editb{to} evolve to have indistinguishable overlap with the original basis states to leading order within $\subsystem{S}$:
\begin{equation}
  \frac{1}{D_E}\Tr[\proj_k(t)\proj_\ell] =\frac{1}{D_S}+o(D_S^{-1}).
    \label{eq:scramblingdef}
\end{equation}
This \editb{is necessary (but not sufficient) for} ``thermalization to infinite temperature''~\cite{Nandkishore, LashkariFastScrambling} and, typically, maximal entanglement~\cite{CanonicalTypicalityPSW, tumulka_CT} at time $t$, i.e., the overlap of the time-evolved states $\hat{\rho}(t) = \proj_k(t)/D_E$ with the observables $\proj_{\ell}$ look like those of the ``infinite temperature'' maximally mixed state $\hat{\rho}_{(\infty)} \equiv \idop/D_S$ in $\mathcal{H}_S$, to leading order. A necessary condition for scrambling in the sense of Eq.~\eqref{eq:scramblingdef} is:
\begin{equation}
    P_S(t) = \frac{1}{D_S} + o(D_S^{-1}).
    \label{eq:ps_scrambling}
\end{equation}
This is further readily shown to be a necessary condition for
scrambling within the subsystem $\subsystem{S}$, of a basis of \textit{pure product states} for $\mathcal{H}$ (see also the supplement~\cite{supplement} for a brief discussion of this~\cite{LashkariFastScrambling} and related notions of scrambling).
From Eqs.~\eqref{eq:dynamicalinequality} and \eqref{eq:ps_scrambling}, we obtain the following \textit{necessary} condition for scrambling at the time $t$ \edit{(usually satisfied within the early time decay of the SFF)}:
\begin{equation}
    K(t) \leq \frac{1}{D_S} + o(D_S^{-1}).
    \label{eq:kt_scrambling}
\end{equation}

Eq.~\eqref{eq:kt_scrambling} is, however, not a \textit{sufficient} condition for scrambling in most bases. For instance, in a basis where $\hat{\mathcal{T}}^t$ has a simple structure e.g. locality, $K(t)$ can even decay to nearly $O(D^{-2}) \ll D_S^{-1}$ by the scrambling timescale~\cite{ShenkerThouless, ChanScrambling, ProsenThermalization}.
However, a typicality result in random matrix theory~\cite{Reimann2016, ChaosComplexityRMT, CotlerHunterJones2, SSS} states that in \textit{almost all} (with respect to the Haar measure~\cite{Mehta, CanonicalTypicalityPSW}) subsystems $\mathcal{H}_S$ and bases $\mathcal{B}_S$ in the Hilbert space of a given system,
\begin{equation}
    \frac{1}{D_E}\Tr[\proj_k(t)\proj_j] = \frac{1+O(D_E^{-1/2})}{D_S}+\left(\delta_{kj}-\frac{1}{D_S}\right)K(t).
    \label{eq:HaarScrambling}
\end{equation}
This means that $K(t) = o(D_S^{-1})$ is a sufficient condition for almost all subsystems of dimension $D_S = o(D)$ in the Hilbert space \edit{(i.e., $N_S/N < 1-\epsilon$)} to scramble at time $t$, which may or may not include a given subsystem of interest.
\edit{Further, such random bases ``almost'' saturate Eq.~\eqref{eq:dynamicalinequality} up to $O(D_S^{-1})$ terms.}

\textit{Fast scrambling in Hamiltonian systems}---In Hamiltonian many-body systems, Eq.~\eqref{eq:kt_scrambling} translates to a constraint on the energy spectrum as follows: if $\dos_D(E) \geq 0$ is the density of states \textit{per level} of the $D$ energy levels of the system (given, at this stage, by a set of delta function spikes that integrates to $1$), then $K(t) = \lvert \tildos_D(t)\rvert^2$, where $\tildos_D(t) = \Tr[\hat{U}(t)]$ is the Fourier transform of $\dos_D(E)$:
\begin{equation}
    \tildos_D(t) \equiv \int \diff E\ \dos_D(E) e^{-iEt}.
    \label{eq:tildosdef}
\end{equation}
This relation \edit{yields} nontrivial constraints on the fastest allowed scrambling time \edit{in Hamiltonian systems} based entirely on the density of states, \edit{where we take $t=\Theta(1)$ to be the timescale of a $\Theta(1)$ decrease in $K(t)$ from $K(0) = 1$}. An important consideration is suggested by a global Haar random (say, CUE~\cite{Haake, Mehta}) \edit{Floquet} unitary $\hat{U}_{\cue}$, which \textit{instantly} scrambles~\cite{ChaosComplexityRMT} an initial state if acting on the system in discrete time steps. However, any \textit{continuous-time} \edit{Hamiltonian} dynamics that reproduces the action of $\hat{U}_{\cue}$ at a time $\tau = \Theta(1)$ [with $e^{-i\hat{H}\tau} = \hat{U}_{\cue}$] typically has \edit{long-time transients} in the SFF, e.g. $K(t) = \sinc^2(\pi t/\tau) \sim t^{-2}$  in the simplest case~\cite{Reimann2016, dynamicalqergodicity}, preventing \textit{sustained} scrambling for a \textit{continuous} length of time until $t_{s,\cue} = \Omega(\sqrt{D_S})$, \edit{even with global interactions.}


\edit{We therefore seek to constrain the time $t_s$, after which a given system can \textit{remain} scrambled up to a long time $T \gg t_s$ beyond all timescales of physical interest (potentially the long $T \sim \exp[\Theta(D)]$ quantum recurrence time~\cite{QuantumRecurrences, BrownSusskind2}). From Eq.~\eqref{eq:kt_scrambling} and $K(t) = \lvert \tildos(t)\rvert^2$ (where we drop the $D$ dependence for convenience) we see that this time is determined by the asymptotic requirement on the density of states:
\begin{equation}
\left\lvert \tildos(t > t_s)\right\rvert^2 \leq \frac{1}{D_S} + o(D_S^{-1}),
\label{eq:ts_dos}
\end{equation}
with $t< T$ implicit.}
\edit{From Eq.~\eqref{eq:ts_dos}, it is clear that the fastest $t_s$ is constrained by what kind of asymptotic behaviors $\tildos(t\to\infty)$ may exist in a ``physical'' system. For instance, power law asymptotics $\lvert\tildos(t\to\infty)\rvert^2 \sim t^{-n}$ implies $t_s \gtrsim D_S^{1/n} = 2^{N_S/n}$, giving a scrambling time that is at least exponentially long in the subsystem size $N_S$. This asymptotic behavior is generally restricted by the non-negativity of a physical density of states $\dos(E) \geq 0$ (as discussed in the supplement~\cite{supplement}), but the quantitative effect of this restriction on the Fourier transform $\tildos(t)$ remains an open mathematical problem.}

\edit{This is because} it is not fully understood mathematically~\cite{FTpositivityConvex, FTpositivity2014} how the non-negativity of $\func(E)$ is reflected in the local or asymptotic behavior of $\tilfunc(t)$. What is known are some necessary~\cite{FTpositivity2014} and sufficient conditions~\cite{FTpositivityConvex} \edit{on $\tildos(t)$ that do not directly address asymptotics~\cite{supplement}}. It is also customary to consider many-body systems with \edit{additional ``physicality conditions'' like} analytic dynamics (in the thermodynamic limit) in many cases~\cite{MSSotocBound, SFFinflectionbound}, 
which enforces $t_s = \omega(1) \gg \Theta(1)$ due to the smooth asymptotic decay of $\tildos(t)$. Given the subjectivity of \edit{imposing} ``physicality'' conditions on the density of states, we instead formulate two exact system-independent statements on fast scrambling \edit{for specific cases of wide relevance}:

\begin{statement}
The Gaussian function $\tilfunc(t) = \exp(-at^2/2)$ is analytic with a positive Fourier transform (also a Gaussian), and consequently quantum systems with fully analytic dynamics that scramble subsystems by any $t_s = \omega(\sqrt{\log D_S})$ do exist, by Eq.~\eqref{eq:HaarScrambling}.
\label{statement:gaussdos}
\end{statement}

\begin{statement}
Due to a Theorem~\cite{FGRexpDecay, SonnerThermodouble, FTcompact} stating that the Fourier transform of a function $\dos(E)$ with one-sided bounded support decays at a slower-than-exponential rate $\lvert \tilfunc(t\to\infty)\rvert \neq O(e^{-\Theta(t)})$, every quantum system with a finite spectral edge~\cite{FGRexpDecay, SonnerThermodouble} as $D\to\infty$ [e.g. $\dos(E < E_0) \to 0$, where $E_0 = O(1)$ over the scale of $\Theta(1)$ variations in $\dos(E)$], \edit{required for a nontrivial (state-independent) ML bound~\cite{ML, flippedML}},  can only scramble subsystems \textit{to infinite temperature} after a time $t_s = \omega(\log D_S)$, by Eq.~\eqref{eq:kt_scrambling}.
\label{statement:sharpdos}
\end{statement}

\edit{\textit{Illustration: Slow scrambling in the SYK model}--- Consider any typical realization of the Majorana SYK-$q$ model~\cite{MaldacenaStanford, KitaevSuh} of fermions on $N$ sites ($2N$ Majoranas, $D=2^N$) subject to random $(q\geq 4)$-body interactions without fermion number conservation (see the supplement~\cite{supplement} for details), which is regarded as ``maximally chaotic'' and a ``fast'' scrambler due to saturating a finite temperature bound on a local-operator measure of ``chaos''~\cite{MSSotocBound}. However, complete scrambling via \textit{entanglement generation} in large subsystems of $N_S$ sites ($D_S = 2^{N_S}$), prepared in Fock states with a maximally mixed (completely uncertain) $E$-component (where $n_{j(k)} \in \lbrace 0,1\rbrace$ is the fermion occupancy number in the $j$-th site):
\begin{equation}
\proj_k = \lvert n_{1(k)},\ldots,n_{N_S(k)}\rangle_S\langle n_{1(k)},\ldots,n_{N_S(k)}\rvert \otimes \idop_E,
\label{eq:SYKinitialstates}
\end{equation}
can proceed ``exponentially'' slowly on account of Eq.~\eqref{eq:dynamicalinequality}.}

\edit{This is because $\dos(E)$ has sharp edges for SYK (similar to Statement~\ref{statement:sharpdos}, albeit at asymptotically large energies)
 leading~\cite{ShenkerThouless,BlackHoleRandomMatrix} to $K(t\to\infty) \sim \Theta(D^{-\alpha}) t^{-3} + O(D^{-1})$, where $\alpha < 1$ ($\alpha \approx 0.66$ for $q=4$, $\alpha \propto q^{-2}$ for large $q$) is determined~\cite{ShenkerThouless} by the zero-temperature entropy~\cite{MaldacenaStanford, SachdevEntropy} $\mathcal{S}_0 = N(1-\tfrac{\alpha}{2})\ln 2$.
 This effect of the edges cannot be ``filtered''~\cite{ShenkerThouless} out of the dynamics due to the completeness of the $\proj_k$~\cite{supplement}.
Thus, from the discussion following Eq.~\eqref{eq:ts_dos}, a subsystem in SYK can remain fully scrambled to any $P_S(t) = O(D_S^{-1})$ (including $D_S^{-1}$) only after an exponentially long time in $N_S$:
\begin{equation}
t_{s, \text{SYK}} = \Omega\left[2^{(N_S-\alpha N)/3}\right], \text{ nontrivial for } N_S > \alpha N.
\label{eq:ts_SYK}
\end{equation}
This $t_{s,\text{SYK}} \gtrsim e^{\Theta(N_S)}$ behavior should be contrasted with, say, systems with a fully Gaussian density of states without sharp edges (e.g.,~\cite{ShenkerThouless}), in which $t_s$ approaches $N_S^{1/2}$ even for finite $N_S/N$ by Statement~\ref{statement:gaussdos}, provided that Eq.~\eqref{eq:HaarScrambling} holds as is typical for systems with sufficiently random energy eigenstates~\cite{Reimann2016}.}

\textit{Conclusions}--- Eq.~\eqref{eq:dynamicalinequality} is a general speed limit on the quantum dynamics of a complete set of states, which remains nontrivial for longer times than \edit{basis-independent versions of} the MT and ML bounds and typically even for asymptotically long times. We showed that it can be used to constrain [Eqs.~\eqref{eq:ps_scrambling} and \eqref{eq:kt_scrambling}] characteristically slow many-body processes, such as the generation of entanglement associated with scrambling or thermalization to infinite temperature. In particular, it enables the problem of the fastest allowed scrambling timescale of a Hamiltonian many-body system to be mapped to a mathematical property of the density of states, irrespective of any interaction structure in the Hamiltonian. \edit{As an illustration, we showed that the SYK model, though ``maximally chaotic'' by certain measures, cannot allow the sustained scrambling (via entanglement generation) of large fermion subsystems up to \textit{exponentially long times in the subsystem size}, due to the sharp edges in its density of states. Another application of  Eq.~\eqref{eq:dynamicalinequality} is to constrain entanglement generation even in finite size systems (e.g., Fig~\ref{fig:SYK}) in terms of the decay of $K(t)$.}

\begin{acknowledgments}
This work was supported by the U.S. Department of Energy, Office of Science, Basic Energy Sciences under Award No. DE-SC0001911 and Simons Foundation. \edit{We thank Michael Winer for useful comments on the SYK zero-temperature entropy}.
\end{acknowledgments}


\bibliography{ScrambledBibliography}

\clearpage
\appendix
\onecolumngrid

\begin{center}
\textbf{\large Exact universal bounds on quantum dynamics and fast scrambling}\vspace{0.2em}

\textbf{\large Supplemental Material}

\vspace{1em}

{\normalsize Amit Vikram$^{1}$\ and Victor Galitski$^{1}$}\vspace{0.2em}

$^{1}$\textit{\small Joint Quantum Institute and Department of Physics, University of Maryland, College Park, MD 20742, USA}

\vspace{1em}
\end{center}

In this supplement, we discuss the derivation of the speed limit in Eq.~\eqref{eq:dynamicalinequality} of the main text as well as its saturation and relation to other speed limits (Sec.~\ref{sec:speedlimit}), some of the notions of scrambling and their relation to Eq.~\eqref{eq:scramblingdef} of the main text (Sec.~\ref{sec:scrambling}), some formal aspects of relating scrambling and the density of states (Sec.~\ref{sec:formal}), and the details of the SYK model as well as how our speed limit leads to a slow sustained scrambling time in this model, in the sense of entanglement generation in large subsystems (Sec.~\ref{sec:SYK}).

\section{Basis-independent quantum speed limit}
\label{sec:speedlimit}
\subsection{Derivation of the inequality \texorpdfstring{$P_S(t) \geq K(t)$}{Ps(t) >= K(t)}}

We have a complete set of orthonormal projectors $\lbrace\proj_k\rbrace$, with
\begin{equation}
    P_S(t) \equiv \frac{1}{D}\sum_{k=1}^{D_S} \Tr\left[\proj_k(t)\proj_k\right].
    \label{eqs:psdef}
\end{equation}
where $\proj_k(t) = \sum_{r=1}^M \kraus{r}{t}\proj_k\krausd{r}{t}$
Let $D_k = \Tr[\proj_k]$ represent the dimensionality of each projector. As $\proj_k$ has $D_k$ eigenvalues equal to $1$, and $(D-D_k)$ eigenvalues equal to $0$, there is an orthonormal set of vectors $\lbrace \lvert k; \ell\rangle\rbrace_{\ell = 1}^{D_k}$ for each $\proj_k$ such that:
\begin{equation}
    \proj_k = \sum_{\ell=1}^{D_k} \lvert k; \ell\rangle \otimes \langle k; \ell\rvert.
    \label{eqs:purestatedef}
\end{equation}
We note that the full set $\bigcup_{k=1}^{D_S}\lbrace \lvert k; \ell\rangle\rbrace_{\ell = 1}^{D_k}$ containing all these vectors forms an orthonormal basis for the Hilbert space $\mathcal{H}$.
Substituting this expression in Eq.~\eqref{eqs:psdef} gives
\begin{align}
    P_S(t) &= \frac{1}{D} \sum_{r=1}^{M} \sum_{k=1}^{D_S} \sum_{\ell,\ell' = 1}^{D_k} \left\lvert \langle k; \ell' \rvert \kraus{r}{t}\lvert k; \ell \rangle\right\rvert^2 \nonumber \\
    &\geq \frac{1}{D} \sum_{r=1}^{M} \sum_{k=1}^{D_S} \sum_{\ell=1}^{D_k} \left\lvert \langle k; \ell \rvert \kraus{r}{t}\lvert k; \ell \rangle\right\rvert^2. \label{eqs:proof1line1}
    \end{align}
In the second line, we have dropped the $\ell \neq \ell'$ terms and retained only the diagonal $\ell = \ell'$ terms (incidentally, such a simple step is also a key element in the proof of the Shnirelman wavefunction ergodicity theorem~\cite{Zelditch}). Now, we consider the contribution to Eq.~\eqref{eqs:proof1line2} from each $\kraus{r}{t}$, for which we get
\begin{align}
  \frac{1}{D}  \sum_{k=1}^{D_S} \sum_{\ell=1}^{D_k} \left\lvert \langle k; \ell \rvert \kraus{r}{t}\lvert k; \ell \rangle\right\rvert^2 &\geq \left[ \frac{1}{D}\sum_{k=1}^{D_S} \sum_{\ell=1}^{D_k} \left\lvert\langle k; \ell \rvert \kraus{r}{t}\lvert k; \ell \rangle\right\rvert\right]^2.
  \label{eqs:proof1line2}
  \end{align}
This is just the inequality $(1/n)\sum_{j=1}^n x_j^2 \geq [(1/n) \sum_{j=1}^n x_j]^2$ with $x_{k;\ell} = \left\lvert \langle k; \ell \rvert \kraus{r}{t}\lvert k; \ell \rangle \right\rvert$ (essentially $\langle x^2\rangle \geq \langle x\rangle^2$, familiar from statistics). The sum on the right hand side can be further constrained (as in the context of ``cyclic aperiodicity'' in Ref.~\cite{dynamicalqergodicity}) using the triangle inequality $\sum_j \lvert y_j\rvert \geq \lvert \sum_j y_j\rvert$ with complex $y_{k;\ell} = \langle k; \ell \rvert \kraus{r}{t}\lvert k; \ell \rangle \in \mathbb{C}$, giving
\begin{equation}
     \frac{1}{D}  \sum_{k=1}^{D_S} \sum_{\ell=1}^{D_k} \left\lvert \langle k; \ell \rvert \kraus{r}{t}\lvert k; \ell \rangle\right\rvert^2 \geq \frac{1}{D^2}\left\lvert \sum_{k=1}^{D_S} \sum_{\ell=1}^{D_k} \langle k; \ell \rvert \kraus{r}{t}\lvert k; \ell \rangle\right\rvert^2 = \frac{1}{D^2}\left\lvert \Tr[\kraus{r}{t}]\right\rvert^2. \label{eqs:proof1line3}
\end{equation}
Combining Eq.~\eqref{eqs:proof1line3} with Eq.~\eqref{eqs:proof1line1}, we get the desired inequality, which is Eq.~\eqref{eq:dynamicalinequality} in the main text: 
\begin{equation}
   P_S(t) \geq  \frac{1}{D^2}\sum_{r=1}^{M} \left\lvert\Tr\left[ \kraus{r}{t}\right]\right\rvert^2 = K(t). \label{eqs:dynamicalinequality}
\end{equation}

\subsection{Condition for saturation}
\label{sec:saturation}

Eq.~\eqref{eqs:dynamicalinequality} is saturated when $P_S(t) = K(t)$. Here, we will identify a sufficient set of conditions for this to happen, which is exhaustive for pure initial states ($D_S = D$) in the case of time-independent unitary evolution (Hamiltonian or Floquet systems). Mixed initial states and other kinds of time evolution present significant additional complications and are outside the scope of this work.

In Hamiltonian and Floquet systems ($M=1$, $\kraus{1}{t} = \hat{U}(t)$), we can expand any pure initial state $\lvert \psi\rangle$ in the (quasi-)energy eigenbasis $\lbrace \lvert E_n\rangle \rbrace_{n=0}^{D-1}$ as $\lvert \psi\rangle = \sum_n c_n \lvert E_n\rangle$. In the special case that the state is equally distributed over all energy eigenstates, $\lvert c_n\rvert^2 = 1/D$, we have
\begin{equation}
\left\lvert \langle \psi\rvert \hat{U}(t)\lvert \psi\rangle\right\rvert^2 = \frac{1}{D^2}\sum_{n,m} e^{i(E_n-E_m)t} = K(t).
\end{equation}
Thus, it follows that any orthonormal basis of states $\lbrace \lvert k\rangle\rbrace_{k=0}^{D-1}$ which are \textit{all} equally distributed over the energy eigenbasis satisfies $P_S(t) = K(t)$ and saturates Eq.~\eqref{eqs:dynamicalinequality}. A simple example of this is a basis of discrete Fourier transforms (DFTs) of the energy eigenstates:
\begin{equation}
\lvert k\rangle = \frac{1}{\sqrt{D}} \sum_{n=0}^{D-1} e^{-2\pi i k n/D} \lvert E_n\rangle.
\end{equation}
The fact that the above speed limit is saturated in this basis of energy DFTs plays a key role in an ergodic classification of quantum dynamical systems by their level statistics~\cite{dynamicalqergodicity}.

A more general basis of pure states that saturate the bound, by the above consideration, can be expressed by transforming the energy eigenbasis using a unitary complex Hadamard matrix $X_{kn}$ (which, by definition~\cite{NielsenChuang}, are unitary matrices such that $\lvert X_{kn}\rvert^2 = 1/D$, ensuring equal distribution in the energy eigenbasis and orthonormality of the transformed basis by virtue of unitarity):
\begin{equation}
\lvert k\rangle = \sum_{n=0}^{D-1} X_{kn} \lvert E_n\rangle.
\label{eqs:HadamardStates}
\end{equation}

The additional complications for other kinds of states and dynamics are as follows: 1. \textit{(for mixed states)} When one has mixed initial states comprised of many pure states, the return probability of the mixed state can involve overlaps of the pure states with each other, and may not saturate the bound even if diagonal in a Hadamard transformation of the energy eigenbasis (where the contribution from the self-overlap of each constituent pure state would already saturate the bound), and 2. \textit{(for non-Hamiltonian/Floquet dynamics)} without a clearly defined basis of energy eigenstates, it is not clear how to identify pure states \textit{within} the Hilbert space whose individual return probability is $K(t)$ as defined in the main text, though this may be possible for the special case of e.g. time-independent non-unitary (i.e. Lindbladian) evolution which admits its own eigenoperators. However, we note that Eq.~\eqref{eq:HaarScrambling} in the main text represents an ``approximate'' saturation of the bound (in an order of magnitude sense) for large subsystems in Haar random bases, as its right hand side is $K(t) + O(D_S^{-1})$, which only differs from $K(t)$ by a small quantity.

\subsection{Tightness compared to other basis-independent speed limits}

Now we will show quite generally, based on the techniques of Ref.~\cite{SonnerThermodouble}, that the speed limit supplied by Eq.~\eqref{eqs:dynamicalinequality} is tighter than other basis-independent speed limit for a complete set of states that may be derived from known speed limits (such as the MT or ML bounds) on individual states, by replacing state-specific properties with those of the spectrum or Kraus operators. Eq.~\eqref{eqs:tighterspeedlimit} is our main result in this regard, while we also state it in the special case of Hamiltonian systems in Eq.~\eqref{eqs:tighterHamiltonianspeedlimit} to aid intuition.

Let $\lvert \psi\rangle$ be a given pure state such that its return probability,
\begin{equation}
P(\lvert \psi\rangle; t) \equiv \sum_{r=1}^{M} \left\lvert \langle \psi\rvert \kraus{r}{t}\lvert \psi\rangle \right\rvert^2,
\end{equation}
satisfies a speed limit set by a function $\speedL$ that depends on the expectation value of powers of the Kraus operators in $\lvert \psi\rangle$:
\begin{equation}
P(\lvert \psi\rangle; t) \geq \speedL\left(\lbrace\langle \psi\rvert [\kraus{r}{t}]^m\lvert \psi\rangle\rbrace ;\ t\right).
\label{eqs:generalspeedlimit}
\end{equation}
For instance, if we are considering the MT bound with a Hamiltonian $\hat{H}$,
\begin{equation}
\speedL\left(\lbrace \langle\psi\rvert e^{-i m\hat{H}t}\lvert\psi\rangle\rbrace;\ t\right) = \cos^2(\Delta_E t),
\end{equation}
where $\Delta_E^2 = \langle \psi\rvert \hat{H}^2\lvert \psi\rangle - \langle \psi\rvert \hat{H}\lvert \psi\rangle^2$ can be extracted from the $\lbrace\langle \psi\rvert e^{-im\hat{H}t}\lvert \psi\rangle\rbrace_{m}$, e.g., through a power series expansion. Of course, one may consider other (e.g., ML) speed limits, in which case $\speedL$ would depend on the expectation values of other functions of $\hat{H}$ in $\lvert \psi\rangle$ in place of the variance $\Delta_E^2$.

With $\mathcal{H}$ being the $D$-dimensional Hilbert space of a system of interest with an orthonormal basis $\lbrace\lvert k\rangle\rbrace_{k=1}^{D}$, we consider a maximally entangled state $\lvert \Psi_{\text{TFD}}\rangle$ in the doubled Hilbert space $\mathcal{H} \otimes \mathcal{H}$, given by:
\begin{equation}
\lvert \Psi_{\text{TFD}}\rangle \equiv \frac{1}{\sqrt{D}}\sum_{k=1}^{D} \lvert k\rangle \otimes \lvert k\rangle.
\end{equation}
This is equivalent to the infinite temperature thermofield double (TFD) state when $\lbrace\lvert k\rangle\rbrace$ is chosen to be the energy eigenbasis in a Hamiltonian system~\cite{SonnerThermodouble}, but we stress that this does not restrict our results to infinite temperatures alone: finite temperature TFD states can always be accounted for through the Kraus operators, e.g., $M=1$, $\kraus{1}{t} = g(\hat{H}) e^{-i\hat{H}t}$ with $g(\hat{H}) \propto e^{-\beta\hat{H}}$ (which has the same effect as choosing a finite temperature TFD state~\cite{SonnerThermodouble} of temperature $\beta^{-1}$ for Hamiltonian evolution). Using the above TFD state, however, has the advantage of allowing us to include non-Hamiltonian e.g. dissipative or time-dependent systems (which don't have obvious energy eigenstates or candidates for finite temperature TFD states) in our considerations.

We assume that time evolution in this doubled Hilbert space acts only on the ``first'' copy with no dynamics in the ``second'', i.e., for density operators $\hat{\rho}$ and $\hat{\sigma}$ in the respective Hilbert spaces (using the notation of Eq. (\maintextEqCPdef) in the main text),
\begin{equation}
\hat{\mathcal{T}}^t_{\mathcal{H}\otimes\mathcal{H}}[\hat{\rho}\otimes\hat{\sigma}] = \hat{\mathcal{T}}^t[\hat{\rho}]\otimes \hat{\sigma}.
\end{equation}
Then, the observation in Ref.~\cite{SonnerThermodouble} that the SFF $K(t)$ for Hamiltonian systems in $\mathcal{H}$ is the return probability of the TFD state also applies to the generalized SFF in Eq.~\eqref{eq:SFFdef} of the main text:
\begin{equation}
K(t) = P(\lvert \Psi_{\text{TFD}}\rangle; t).
\label{eqs:genSFFTFD}
\end{equation}
Here, it is important to re-emphasize that $K(t)$ is a \textit{state-independent} quantity  (being just a double trace of the time evolution operator) of the original system in $\mathcal{H}$, while $P(\lvert \Psi_{\text{TFD}}\rangle; t)$ is a naturally state-dependent quantity in the doubled Hilbert space $\mathcal{H}\otimes \mathcal{H}$. Equating the two does not affect the state-independence of the former; instead, the equality states that the dynamics of the specific doubled-Hilbert-space state $\lvert \Psi_{\text{TFD}}\rangle \in \mathcal{H} \otimes \mathcal{H}$ captures \textit{state-independent} aspects of the dynamics within one copy of the Hilbert space $\mathcal{H}$.

Using Eq.~\eqref{eqs:genSFFTFD} in \eqref{eqs:generalspeedlimit}, and noting that the TFD state captures only the \textit{state-independent} traces of the Kraus operators within the Hilbert space $\mathcal{H}$,
\begin{equation}
\langle \Psi_{\text{TFD}}\rvert [\kraus{r}{t}]^m\lvert \Psi_{\text{TFD}}\rangle = D^{-1}\Tr\left[[\kraus{r}{t}]^m\right],
\end{equation}
immediately gives our desired relation:
\begin{equation}
K(t) \geq \speedL\left(\left\lbrace D^{-1}\Tr\left[[\kraus{r}{t}]^m\right]\right\rbrace;\ t\right).
\label{eqs:Ktspeedlimit}
\end{equation}
We note that it is not essential to deal with a doubled Hilbert space to derive this inequality in certain cases. For Hamiltonian (or Floquet) systems, this can also be derived from $P(\lvert k\rangle; t) = K(t)$ for the states of Eq.~\eqref{eqs:HadamardStates} or as a special case of a geometric bound on the growth of errors in unitary approximations of quantum dynamical systems~\cite{dynamicalqergodicity}.
But the thermofield double appears to provide a more useful general technique for non-Hamiltonian and non-Floquet evolution where one has no further information about the structure of the Kraus operators.

From $P_S(t) \geq K(t)$, it further follows that
\begin{equation}
P_S(t) \geq K(t) \geq \speedL\left(\left\lbrace D^{-1}\Tr\left[[\kraus{r}{t}]^m\right]\right\rbrace ;\ t\right).
\label{eqs:tighterspeedlimit}
\end{equation}
Let us review the physical content of Eq.~\eqref{eqs:tighterspeedlimit}. It states that provided one has a formula for a \textit{pure state} quantum speed limit $L$ in terms of the expectation value of powers of the Kraus operators in a specific state (which may be generalizations of MT, ML or tighter formulations), one can obtain a basis-independent speed limit on $P_S(t)$ (which we recall is the \textit{mean return probability} of a complete set of states, rather than an individual state) from $L$, but now depending on the state-independent traces of powers of the Kraus operators. However, $K(t)$ also satisfies this speed limit [Eq.~\eqref{eqs:Ktspeedlimit}], and Eq.~\eqref{eqs:dynamicalinequality} therefore sets a tighter basis-independent bound on $P_S(t)$ than $L$.

Let us see how this works for the Hamiltonian case to build intuition. Here, the traces $\lbrace \Tr[e^{-i m\hat{H}t}]\rbrace_{m}$ can be transformed a power series expansion (for convenience) directly to the state-independent moments of the eigenvalues of the Hamiltonian:
\begin{equation}
\langle E^m\rangle \equiv D^{-1} \Tr[\hat{H}^m].
\end{equation}
These are the statistical properties of the energy spectrum itself, rather than any particular state within $\mathcal{H}$. In this case, Eq.~\eqref{eqs:tighterspeedlimit} becomes:
\begin{equation}
P_S(t) \geq K(t) \geq \speedL\left(\left\lbrace \langle E^m\rangle \right\rbrace;\ t\right).
\label{eqs:tighterHamiltonianspeedlimit}
\end{equation}
Thus, $L$ yields a state-independent speed limit on $P_S(t)$ as determined by properties of the energy spectrum itself, but the speed limit set by $K(t)$ is always tighter. When $L = \cos^2(\Delta_E t)$ corresponding to the MT bound, we replace $\Delta_E^2$ (the variance of a state) with $\sigma_E^2$ (the variance of the energy spectrum itself) to get Eq.~\eqref{eq:sffcosbound} of the main text.

Finally, as already stated in the main text, we emphasize that $K(t)$ does not set any direct speed limit on \textit{individual} states, and is not always tighter than state or basis-\textit{dependent} speed limits for specific states. For instance, any state in the energy eigenbasis of a Hamiltonian system has $\Delta_E = 0$ and $P(\lvert E_n\rangle; t)=1=\cos(0 t)$, saturating the MT bound but not Eq.~\eqref{eqs:dynamicalinequality} for the basis states; however, almost all other states in the Hilbert space $\mathcal{H}$ with nontrivial dynamics violate this state-specific MT speed limit set by $\cos(0t) = 1$, as $P(t) \leq 1$ in general.

Thus, what Eq.~\eqref{eqs:tighterspeedlimit} shows is that Eq.~\eqref{eqs:dynamicalinequality} essentially remains the tightest possible state-independent speed limit that applies to \textit{all complete sets of states} in the Hilbert space, as any other speed limit $L$ for return probabilities also sets a state-independent bound [Eq.~\eqref{eqs:Ktspeedlimit}] on the SFF. We note that this makes it especially suited to address questions such as the fastest \textit{allowed} scrambling time for any basis in the system, which is the key application we focus on in this work.

\section{Types of scrambling and thermalization}
\label{sec:scrambling}

Here, we discuss some of the different physically interesting types of scrambling and their relation to thermalization as well as the discussion in the main text.

\subsection{The scrambling of added degrees of freedom to infinite temperature}
We begin with the type of scrambling directly defined in the main text, in the setting of a Hilbert space $\mathcal{H} = \mathcal{H}_S \otimes \mathcal{H}_E$ where the subsystem $\mathcal{H}_S$ is $D_S$-dimensional, and the component $\mathcal{H}_E$ of the system external to the subsystem is $D_E$-dimensional. According to Eq.~\eqref{eq:scramblingdef} of the main text, scrambling at time $t$ corresponds to:
\begin{equation}
  \frac{1}{D_E}\Tr[\proj_k(t)\proj_\ell] =\frac{1}{D_S}+o(D_S^{-1}),
    \label{eqs:scramblingdef}
\end{equation}
where $\proj_k/D_E$ are a complete set of ($D_E$-dimensional) orthonormal initial states, that have the form $\proj_k = \lvert k\rangle_S\langle k\rvert \otimes \idop_E$, where the $\lvert k\rangle_S$ form an orthonormal basis in $\mathcal{H}_S$. We noted in the main text that this corresponds to thermalization to infinite temperature, which essentially means that $\proj_k(t)/D_E$ ``looks'' like the maximally mixed (or infinite temperature) thermal state $\hat{\rho}_{S}(\infty) = \idop_S/D_S$ within $\mathcal{H}_S$ (see also the discussion in Ref.~\cite{Nandkishore}); correspondingly, Eq.~\eqref{eqs:scramblingdef} also follows by requiring the overlaps of $\proj_k(t)/D_E$ with all $\proj_\ell$ to look like those of $\hat{\rho}_S(\infty)$ to leading order. We note, however, that Eq.~\eqref{eqs:scramblingdef} is slightly more general, and may be satisfied if $\proj_k(t)/D_E$ is not close to a maximally mixed state within $\mathcal{H}_S$, e.g. even if it reduces to a pure state that is completely delocalized over the original basis. In this sense, the above notion of scrambling is only a \textit{necessary} condition for the generation of maximal entanglement within $\mathcal{H}_S$.

Let us consider a physical procedure where the above form of scrambling is relevant. As the initial state $\hat{\rho}_k(0) = \proj_k/D_E = \lvert k\rangle_S\langle k\rvert \otimes (\idop_E/D_E)$ is pure in the subsystem $\mathcal{H}_S$ and maximally mixed in the external component $\mathcal{H}_E$, this corresponds to the following situation: say we start with a many-body system with Hilbert space $\mathcal{H}_E$, which has already thermalized to infinite temperature via maximal entanglement with some larger space $\mathcal{H}_R$. At time $t=0$, we couple it to the subsystem $\mathcal{H}_S$ comprised of a set of qubits in a product state $\lvert k\rangle$, and allow the combined system $\mathcal{H} = \mathcal{H}_S \otimes \mathcal{H}_E$ to evolve (without any interaction with $\mathcal{H}_R$); Eq.~\eqref{eqs:scramblingdef} then corresponds directly to the scrambling of this $\hat{\rho}_k(t)$ for all initial product states of the \textit{added} subsystem.

This type of scrambling of ``added degrees of freedom'' $\mathcal{H}_S$ to an already scrambled system $\mathcal{H}_E$, the latter being entangled with an additional system $\mathcal{H}_R$, is of the most direct relevance to the Hayden-Preskill protocol and the black hole information problem~\cite{HaydenPreskill, SekinoSusskind, LashkariFastScrambling}. In the restricted case of $D_S = \Theta(1)$, the bounds offered by Eq.~\eqref{eqs:dynamicalinequality} only constrain scrambling to occur slower than some $\Theta(1)$ time (over which $K(t)$ changes by a $\Theta(1)$ amount), which is much shorter than the associated scrambling time of e.g. systems with $[k=O(1)]$-local interactions that is expected~\cite{HaydenPreskill, SekinoSusskind, LashkariFastScrambling, BentsenGuLucasScrambling, LucasEntanglementVsOTOC} to be $t=\Omega(\log\log D)$ (especially in large subsystems). This is largely because Eq.~\eqref{eqs:dynamicalinequality} allows us to consider only the return probabilities \textit{within} the added $\mathcal{H}_S$ (i.e. overlaps of $\proj_k(t)$ with the original $\proj_k$), as opposed to a fuller consideration of the scrambling of these initial states within $\mathcal{H}_S$ over arbitrary subsystems much larger than $\mathcal{H}_S$. We note that bounds corresponding to asymptotically large times $t = \omega(1) \gg \Theta(1)$ are still directly obtained using $P_S(t)$ if one considers adding much larger subsystems with $D_S = \omega(1)$ in this procedure. Additionally, our bound allows us to nontrivially consider arbitrarily large subsystems in a related case of more immediate physical interest as discussed below: the scrambling of initial states that are product states in the \textit{full} system.

\subsection{The scrambling of pure product states to infinite temperature}
\label{sec:pureproductstates}
Now, given a set of $\lbrace \proj_k\rbrace$, consider initial states corresponding to any basis of their constituent pure states, e.g., $\proj_{k;\ell} \equiv \lvert k;\ell\rangle\langle k;\ell\rvert$ (see also Eq.~\eqref{eqs:purestatedef}). Then, we have from Eqs.~\eqref{eqs:psdef} and \eqref{eqs:purestatedef}:
\begin{equation}
    P_S(t) = \frac{1}{D}\sum_{k=1}^{D_S} \sum_{\ell=1}^{D_E} \Tr[\proj_{k; \ell}(t)\proj_k].
    \label{eqs:purestateps}
\end{equation}
Each term in Eq.~\eqref{eqs:purestateps} corresponds to the following situation: prepare the pure state $\lvert k;\ell\rangle \in \mathcal{H}$, and consider its overlap with its restriction $\lvert k\rangle_S \in \mathcal{H}_S$. For instance, the initial states $\proj_{k;\ell}$ could be product states like $\lvert 0_1 1_2 0_3 1_4\rangle$ in a computational basis of $4$ qubits, while $P_S(t)$ measures their mean overlap with e.g. the corresponding restricted states $\lvert 0_1 1_2\rangle_S \in \mathcal{H}_S$ in a $2$-qubit subsystem (of, say, the first $2$ qubits). Here, we are free to choose $\mathcal{H}_S$ to be whichever subsystem of qubits we want (in particular, of any size), and the scrambling of all such product states in a computational basis within any chosen subsystem of dimension $D_S$ at time $t$ requires that $P_S(t) = D_S^{-1} + o(D_S^{-1})$. Thus, Eq.~\eqref{eq:ps_scrambling} of the main text is again a necessary condition for this more accessible form of scrambling.

Such bounds would also apply to a version of the Hayden-Preskill~\cite{HaydenPreskill} protocol in place of the discussion in the preceding subsection, if it can be extended to a basis of pure initial states in the full $\mathcal{H}_S \otimes \mathcal{H}_E \otimes \mathcal{H}_R$ Hilbert space to eventually appear scrambled within arbitrarily large subsystems of the smaller $\mathcal{H}_S \otimes \mathcal{H}_E$ interacting subspace (we recall that $\mathcal{H}_R$ does not participate in dynamics).

\subsection{Thermalization to finite temperature}

Similar to the infinite temperature case, we may consider a state to have thermalized to a finite temperature $\beta^{-1}$ if it appears indistinguishable from a given finite temperature density matrix $\hat{\rho}_{(\beta^{-1})}$ within a subsystem. Such density matrices are usually obtained as the reduction of a microcanonical state,
\begin{equation}
    \hat{\rho}_{E_\beta, \Delta E} \propto \sum_{E_n \in [E_\beta,E_\beta+\Delta E]} \lvert E_n\rangle\langle E_n\rvert
    \label{eqs:finitetemp}
\end{equation}
supported on a narrow range of energies $[E_\beta,E_\beta+\Delta E]$, to the subsystem~\cite{DAlessio2016,deutsch2018eth,subETH}. The simplest way to apply the bound of Eq.~\eqref{eqs:dynamicalinequality} to this case is to consider a complete set of states within the smaller Hilbert space spanned by the energy eigenstates within $[E_\beta, E_\beta+\Delta E]$.

A complication arises when one considers subsystems in place of an arbitrary basis of states within the microcanonical window. A complete set of states within the subsystem is necessarily supported on the \textit{entire} Hilbert space (by virtue of its completeness) rather than a microcanonical window, and its dynamics consequently involves the full range of energies available in the system --- thereby corresponding to infinite temperature thermalization, if one were to require any form of scrambling with respect to such states. It therefore appears that, in the case of finite temperature thermalization, one cannot presume to have complete control over the state of subsystems to the extent of preparing any pure state in a given basis; subsystem states must instead be ``coarse grained'' in some way to reflect only those states of the subsystem supported within $[E_\beta, E_\beta+\Delta E]$.

Alternatively, it is customary to consider ``regularized'' time evolution operators such as $g(\hat{H})e^{-i\hat{H}t}$ as accounting for such finite temperature effects, e.g. when $g(\hat{H})$ is significantly supported~\cite{Reimann2016} only within $[E_\beta, E_\beta+\Delta E]$ or sometimes~\cite{BlackHoleRandomMatrix, ShenkerThouless} with $g(\hat{H}) \propto e^{-\beta\hat{H}}$. For such a prescription to be useful in our context, in a way that allows applying Eq.~\eqref{eqs:dynamicalinequality} to a physically meaningful process of thermalization to finite temperature, the choice of $g(\hat{H})$ will have to be justified by an appropriate restriction on the kind of initial states that one may prepare within the subsystem. \edit{In either case, our inequality continues to constrain the thermalization of such coarse-grained states states or under regularized time evolution operators to the appropriate finite temperature value of $P_S(t)$.}

\subsection{Entanglement vs operator spreading}
\label{sec:entVsOp}

In addition to the form of scrambling represented by Eq.~\eqref{eqs:scramblingdef}, which we noted to be a necessary condition for the generation of a maximal degree of entanglement, another commonly used notion of scrambling is that of the saturation of out-of-time order correlators (OTOCs) to $\Theta(1)$ values, which measures the extent of ``operator spreading'' in a many-body system. While this can often take asymptotically long times when the growth of OTOCs is slow, the criterion of saturation to $\Theta(1)$ values does not appear to have sufficient resolution to accurately track the generation of entanglement over a large number of degrees of freedom (e.g. it seems analogous to requiring $P_S(t) \ll 1$ in our case, as opposed to the more precise $P_S(t) = D_S^{-1}+o(D_S^{-1})$). It has also been noted that the saturation of OTOCs can occur in $\Theta(1)$ times even in $k$-local systems, much faster than the time it takes for entanglement to be generated~\cite{BentsenGuLucasScrambling, LucasEntanglementVsOTOC}. For this reason, Eq.~\eqref{eqs:scramblingdef} appears to be the most relevant definition of scrambling for the present study, even independent of its natural connection with Eq.~\eqref{eqs:dynamicalinequality}.

\section{Fast Scrambling and Fourier transforms: Formal statement}
\label{sec:formal}
\edit{This section formally states the connection between fast scrambling and properties of Fourier transforms noted in the main text in a certain time regime, and may be skipped by a reader interested more in the physical implications of the speed limit.}

The question of \textit{fast scrambling}, in a form that takes the discussion of ``sustained'' scrambling in the main text into account, considers whether it is possible for thermodynamically large subsystems of size $D_S$ (implicitly given in terms of $D$, e.g. $D_S = D^{1/4}$) to become and remain scrambled after a given time $t_s = \Theta[\invscrtimef(D_S)]$ up to a long time $T > t_s$ (potentially the long $T \sim \exp[\Theta(D)]$ quantum recurrence time~\cite{QuantumRecurrences, BrownSusskind2}).
Here, $\scrtimef(x) > 0$ is a given monotonically increasing function of $x > 0$, with inverse $\invscrtimef(\scrtimef(x))=x$.
By Eq.~\eqref{eq:kt_scrambling} of the main text,
this requires (with $t<T$ implicit):
\begin{equation}
    K\left(t > t_s =  \Theta[\invscrtimef(D_S)]\right) \leq \frac{1}{D_S} + o(D_S^{-1}).
    \label{eq:tmin}
\end{equation}
Further, Eq.~\eqref{eq:HaarScrambling} of the main text
implies that
\begin{equation}
    K\left(t > t_s =  \Theta[\invscrtimef(D_S)]\right) = o(D_S^{-1})
    \label{eq:trmt}
\end{equation}
is a sufficient condition for fast scrambling by $t_s = \Theta[\invscrtimef(D_S)]$ in \textit{some} subsystem of dimension $D_S$ in the Hilbert space.
We note that the $\Theta(1)$ time scale is set here by that of significant $\Theta(1)$ variations in the SFF (i.e. the time scale over which $K(t)$ decreases from $K(0) = 1$ by a finite amount). The fastest scrambling system is one that satisfies the above criterion with the fastest growing $\scrtimef(x)$, or equivalently, the slowest growing $t_s$ (say, among a set of systems subject to certain ``physicality'' conditions on the SFF). While it has been conjectured and argued that $t_s = \Omega(\log\log D)$ --- \edit{with a suitable prefactor (e.g. inverse temperature) that sets an $O(1)$ time scale} --- in systems with $[k=O(1)]$-local interactions~\cite{SekinoSusskind, LashkariFastScrambling, BentsenGuLucasScrambling}, we would like to explore this question in a general setting.
In our approach, on setting $K(t) = \lvert \tildos_D(t)\rvert^2$ in Eqs.~\eqref{eq:tmin} and \eqref{eq:trmt} \edit{(where $\dos_D(t)$ is the Fourier transform of the density of states $\dos(E)$, as in Eq.~\eqref{eq:tildosdef} of the main text)}, the problem of fast scrambling \edit{is determined by} the $D_S, D\to\infty$ asymptotic behavior of $\lvert \tildos_D(\Omega[\invscrtimef(D_S)])\rvert^2$.

To simplify the present analysis, we will focus on an early-time regime $\lvert t\rvert < t_{D}$ where we assume negligible \textit{explicit} $D$-dependence \edit{(to avoid complications from having $2$ independent asymptotic parameters $D_S$ and $D$ in this initial study)}:
\begin{equation}
    \tildos_D\left(t : \lvert t\rvert < t_{D}\vphantom{\tildos}\right) = \tildos(t) + o\left(\tildos(t)\right),
    \label{eq:dmirrelevantdef}
\end{equation}
for a $D$-independent function $\tildos(t)$. By Eq.~\eqref{eq:tmin}, $\tildos_D(t)$ can be replaced with $\tildos(t)$ in the fast scrambling problem for subsystems of dimension $D_S = o(\maxds)$, where $\maxds \equiv 1/K(t_{D})$; we will correspondingly call $\tildos(t)$ the ``smoothened'' form of $\tildos_D(t)$. It is further convenient to smoothen out the finer structure of $\dos_D(E)$ (by e.g. convolving it with a Gaussian of width $\sigma \sim 2\pi/t_{D}$), so that it essentially equals the $D$-independent Fourier transform $\dos(E) \geq 0$ of $\tildos(t)$ \edit{(as such a convolution suppresses the $\lvert t\rvert > t_D$ oscillations of $\tildos_D(t)$)}.
Several physical systems can even have $\maxds \sim D$ (corresponding to effective $D$-independence at early times when $K(t) \gg D^{-1}$), such as typical realizations of the random matrix ensembles~\cite{Haake, Mehta} and the (exact or numerically apparent) behavior of some of the many-body systems considered in Refs.~\cite{ShenkerThouless, ExpRamp1, ExpRamp2, MBLSFF}, where this smoothening can be made rigorous through ensemble averaging \edit{(noting that ``each'' energy level typically follows a smooth distribution, though not necessarily Gaussian, over infinitely many ensemble realizations).  As another example, the SYK-$(q \geq 4)$ model~\cite{MaldacenaStanford, KitaevSuh, BlackHoleRandomMatrix} considered in the main text and Sec. \ref{sec:SYK} appears to have $D_m \sim D^{\alpha}$ with $\alpha < 1$ (see Eq.~\eqref{eqs:SYKsff}), due to which an extension of these formal considerations that accounts for $D$-dependence is necessary to directly tackle Eq.~\eqref{eqs:SYKscramblingtime} applying to larger subsystems.}

With this simplification (which is essentially an argument to use a smooth $D$-independent density of states),
we can formally state the combination of Eqs.~\eqref{eq:tmin} and \eqref{eq:trmt}, using 
\eqref{eq:dmirrelevantdef} and the fact that $\lvert \tildos(0)\rvert = 1$ from the properties of $K(t)$, as a proposition relating the question of whether the scrambling of subsystems can occur before a given time scale $\invscrtimef(D_S)$ to mathematical necessary conditions on the asymptotic decay of a (sufficiently well-behaved) function to ensure a nonnegative Fourier transform:

\begin{proposition}[\textbf{Fast scrambling and Fourier transform nonnegativity}]
Let $\funcset$ be the set of functions $\tilfunc: \mathbb{R} \to \mathbb{C}$ normalizable to $\lvert \tilfunc(0)\rvert = 1$ with a non-negative real-valued Fourier transform $\func(E) \geq 0$, and $\funcsubset \subseteq \funcset$ be any subset of these functions satisfying as yet unspecified ``physicality'' conditions. \textbf{If} every such ``physical'' $\tilfunc(t) \in \funcsubset$ necessarily has a slow asymptotic decay satisfying:
\begin{equation}
  \lvert \tilfunc(t \to \infty)\rvert^2 \neq  o\left[\frac{1}{\scrtimef\left(\Theta\left(t\right)\right)}\right],
   \label{eq:FourierPositivityAsymptotics}
\end{equation}
where $\scrtimef(x > 0) > 0$ is a given monotonically increasing function, \textbf{then} any quantum system whose dynamics is subject to the same physicality conditions [i.e. with smoothened $\tildos(t) \in \funcsubset$ and the corresponding smoothened density of states $\dos(E)$] can scramble subsystems of dimension $D_S = o(\maxds)$ only after a time $t_s = \Omega(\invscrtimef(D_S))$ [by Eq.~\eqref{eq:tmin}].
\textbf{Further}, a quantum system with $\tildos(t) \in \funcsubset$ that scrambles subsystems within $t_s = O(\invscrtimef(D_S))$ is guaranteed to exist if there is at least one function $\tilfunc_1(t) \in \funcsubset$ that decays faster than in Eq.~\eqref{eq:FourierPositivityAsymptotics}, i.e. $\lvert \tilfunc_1(t \to \infty)\rvert^2 = o\left[1/\scrtimef\left(\Theta\left(x\right)\right)\right]$ [from Eq.~\eqref{eq:trmt}].
\label{prop:FSFTP}
\end{proposition}

This proposition is our main \edit{formal statement} as far as Hamiltonian fast scrambling is concerned, setting lower limits on the scrambling time \edit{within the regime of effective $D$-independence in $\tildos_D(t)$}. The subset of functions $\funcsubset$ \edit{left unspecified above} may be chosen according to certain ``physicality'' conditions on the spectrum, e.g. the analyticity of $K(t)$ in the $\lvert t\rvert < t_{D}$ regime. 
What makes the resulting problem nontrivial is that one has to simultaneously satisfy $\dos(E) \geq 0$ \textit{and} the physicality conditions.

\edit{As noted in the main text, it is not known~\cite{FTpositivityConvex, FTpositivity2014} how such conditions determine the asymptotic behavior of Fourier transforms.} What is known are certain necessary conditions~\cite{FTpositivity2014} such as maximality $\tilfunc(0) \geq \tilfunc(t)$ (automatically satisfied in our case due to the properties of $K(t)$) and concavity $\tilfunc''(t=0) < 0$ at the origin, and sufficient conditions~\cite{FTpositivityConvex} such as convexity $\tilfunc''(t>0) > 0$ for real-valued and symmetric $\tilfunc(t) = \tilfunc(-t)$. Such convex functions can be made to decay as fast as desired (corresponding to $t_s$ as small as desired in physical systems): an extreme limiting case with $t_s = 1$ is $\tilfunc(t) = \max\lbrace 1-\lvert t\rvert,0\rbrace$, with $\func(E) = \dos(E) = \sinc^2(\pi E) \geq 0$. However, they are non-analytic at $t=0$, possessing an infinite energy variance $\sigma_E^2 = -K''(0)/2$; consequently, the scrambling time \textit{relative} to the time scale set by $\sigma_E$ is infinite even in this case, i.e., $\sigma_E t\to\infty$. \edit{We also note that the (state-independent) ML bound for $P_S(t)$ in this case would be trivial, $P_S(t\neq 0) \geq 0$, as $\dos(E) = \sinc^2(\pi E)$ admits a spectrum of infinite width, without a finite ground or ceiling state in the $D\to\infty$ limit (alternatively, as noted in Statement 2 of the main text, a nontrivial ML bound would automatically imply a higher-than-logarithmic scrambling time in $D_S$, and cannot lead to $t_s = 1$).}

\section{Example: Slow scrambling via entanglement generation in the SYK model}
\label{sec:SYK}

Here, we provide additional details pertaining to the scrambling of information in the Sachdev-Ye-Kitaev (SYK) model~\cite{MaldacenaStanford, KitaevSuh} discussed in the main text (see Figs.~\ref{fig:SYKdosSFF}, \ref{fig:SYKscrambling} below for a quick numerical overview of the main takeaways). The conventional form of the SYK model consists of Majorana fermions $\hat{\chi}_j$ on $2N$ sites with random all-to-all $4$-body interactions with antisymmetric coefficients $J_{jk\ell m}$, where each independent $J_{jk\ell m}$ is randomly chosen from a Gaussian distribution with variance $\langle J_{jk \ell m}^2\rangle = 3!J^2/N^3$; the Hamiltonian is:
\begin{equation}
\hat{H} = \sum_{1 \leq j < k < \ell < m \leq 2N} J_{jk\ell m} \hat{\chi}_j \hat{\chi}_k \hat{\chi}_\ell \hat{\chi}_m.
\label{eqs:defSYK4}
\end{equation}
A slight generalization is the SYK-$q$ model with $q$-body random interactions with $J_{j_1\ldots j_q}$ chosen from independent Gaussians of variance $(q-1)!J^2/N^{q-1}$, given by
\begin{equation}
\hat{H} = -i^{q/2}\sum_{1 \leq j_1 < \ldots < j_q \leq 2N} J_{j_1 \ldots j_q} \hat{\chi}_{j_1} \ldots \hat{\chi}_{j_q}.
\label{eqs:defSYK}
\end{equation}
Both in the main text and in the supplement, we are only concerned with a single representative realization of the interaction coefficients $J_{{j_1}\ldots{j_q}}$ chosen from the above distribution with $J=1$, and do not consider any form of ensemble averaging. A key feature of the SYK model is that its regularized OTOCs (essentially, combinations 4-point correlation functions of certain regularized single-site operators) grow exponentially fast at the maximum allowed rate~\cite{MSSotocBound}, due to which the model is considered to be ``maximally chaotic'' and a ``fast scrambler''. While this statement remains true if one considers scrambling in the sense of growth of OTOCs, the scrambling of information via entanglement generation over large subsystems can be quite different as noted in Sec.~\ref{sec:entVsOp} (see also Ref.~\cite{LucasEntanglementVsOTOC}). In particular, the criterion for a rapid growth of OTOCs merely to $O(1)$ values may not have sufficient resolution to capture microscopic deviations from scrambling at a resolution comparable to $D_S^{-1}$ for large subsystems.

As discussed in the main text, with details provided below, the density of states of the SYK model is such that the sustained scrambling time of large subsystems is at least exponentially long in the subsystem size, while a generic strongly interacting many-body system with a Gaussian density of states and sufficiently ``random'' eigenstates would scramble much faster as a square root of the subsystem size. Qualitatively, this is because of slowly decaying oscillations in the SFF $K(t)$ for this model caused by sharp edges in its density of states, which through the inequality Eq.~\eqref{eqs:dynamicalinequality} forces the mean return probability of initial states in such subsystems to ``remember'' the initial state at least until these oscillations decay to a sufficiently small value.

\subsection{Defining subsystems in the SYK model}
We would like to consider information scrambling in subsystems of this model. To define suitable subsystems, it is convenient to transform from the $2N$ Majoranas to complex fermions on $N$ sites, given by:
\begin{align}
\hat{c}_j = \frac{1}{\sqrt{2}}\left(\hat{\chi}_{2j}-i\hat{\chi}_{2j+1}\right), \label{eqs:cdef}\\
\hat{c}_j^\dagger = \frac{1}{\sqrt{2}}\left(\hat{\chi}_{2j}+i\hat{\chi}_{2j+1}\right) \label{eqs:cdaggerdef}.
\end{align}
These admit a vacuum state $\lvert 0\rangle$ such that $\hat{c}_j\lvert 0\rangle = 0$ for all $j$, and the Hilbert space $\mathcal{H}$ is spanned by orthonormal Fock states:
\begin{equation}
\lvert n_1, n_2,\ldots,n_N\rangle \equiv (\hat{c}_N^\dagger)^{n_N}\ldots (\hat{c}_2^\dagger)^{n_2} (\hat{c}_1^\dagger)^{n_1}\lvert 0\rangle,\ \text{for } n_j \in \lbrace 0,1\rbrace,
\label{eqs:Fockgendef}
\end{equation}
where we have chosen a specific ordering of fermion operators for definiteness, and reordering them only changes the overall sign of each state. It is important to emphasize that the Hamiltonian $\hat{H}$ in Eq.~\eqref{eqs:defSYK} does not conserve the particle number in terms of these fermions, and is quite unlike the closely related complex SYK model~\cite{GKSTcomplexSYK} which does conserve fermion number. In particular, the nonconservation of fermion number allows the scrambling of subsystems comparable to infinite temperature thermalization as considered in the main text.

Now we can formally factorize the Hilbert space, exactly like a system of qubits, into a subsystem $\mathcal{H}_S$ of dimension $D_S = 2^{N_S}$ representing $N_S$ sites, for instance, $\lbrace 1,\ldots,N_S\rbrace$, and a factor $\mathcal{H}_E$ of dimension $D_E = 2^{N_E}$ representing the remaining $N_E = N-N_S$ sites:
\begin{equation}
\lvert n_1, n_2,\ldots,n_N\rangle = \lvert n_1,\ldots,n_{N_S}\rangle_S \otimes \lvert n_{N_S+1},\ldots,n_{N}\rangle_E.
\label{eqs:subsysdef}
\end{equation}
We also recall that while the fermion ladder operators $\hat{c}_j$ can not be factorized in this manner between sites due to their mutual anticommutation relations, the fermion number operators $\hat{n}_j = \hat{c}_j^\dagger \hat{c}_j$ are mutually commuting and do factorize into those with $j \leq N_S$ acting on $\mathcal{H}_S$ alone, and the remaining acting on $\mathcal{H}_E$ alone.

Now, we can consider $P_S(t)$ for a set of $D_S$ initial states $\proj_k$ corresponding to the Fock basis states within $\mathcal{H}_S$, with the state in $\mathcal{H}_E$ being maximally mixed (representing complete uncertainty in the configuration of $\mathcal{H}_E$):
\begin{equation}
\proj_k = \lvert n_{1(k)},\ldots,n_{N_S(k)}\rangle_S \langle n_{1(k)},\ldots,n_{N_S(k)}\rvert \otimes \idop_E.
\label{eqs:projdef}
\end{equation}

Eq.~\eqref{eqs:dynamicalinequality} states that $P_S(t) \geq K(t)$, and therefore, as discussed in the main text, the scrambling time of the SYK model depends on the asymptotic structure of $K(t\to\infty)$, which we describe next.

\subsection{Power-law decay of the SFF}

\begin{figure}[!hbt]
\subfloat[][Density of states (per level) $\dos(E)$, coarse grained over bins of width $0.05$ in energy units where $J=1$, for the specific realization of the SYK model considered here, showing the sharp edges.]{\includegraphics[width=0.45\columnwidth]{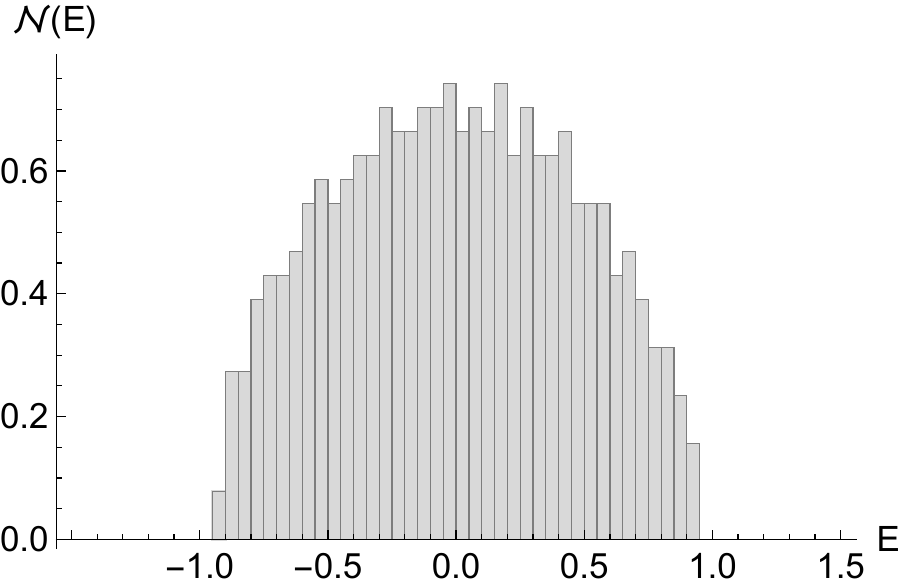} \label{fig:SYKdos}} \qquad
\subfloat[][Log-log plot of SFF $K(t)$ compared with a $t^{-3}$ decay of early oscillations after the first decay, verifying Eq.~\eqref{eqs:SYKsff}. The late-time fluctuations are the ramp and (barely visible) plateau which remain comparable to $O(D^{-1})$; these would appear much more negligible in the $D\to\infty$ limit, and have no direct effect on the scrambling of subsystems of dimension $D_S \ll D$.]{\includegraphics[width=0.45\columnwidth]{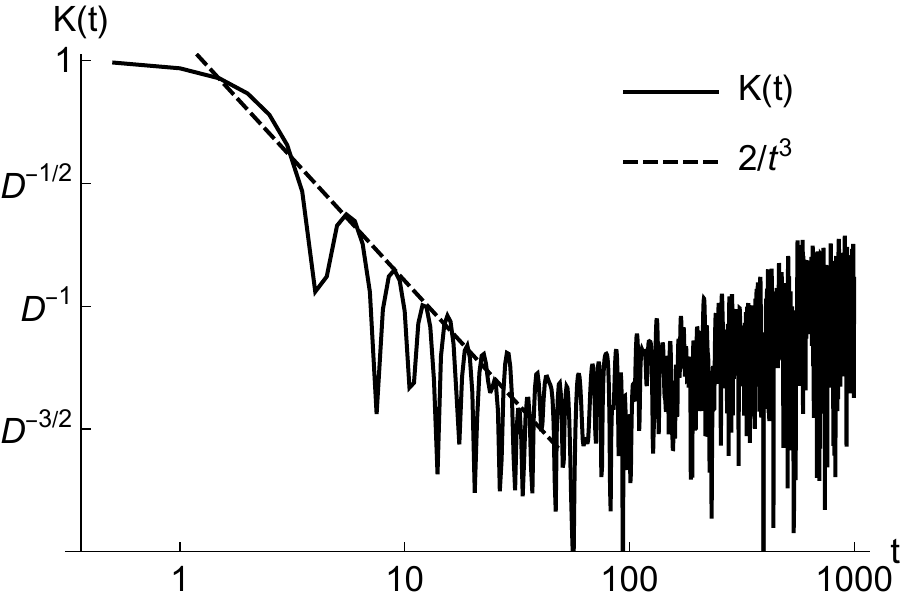} \label{fig:SYKSFFt3}}
\caption{Density of states and SFF for the same realization of the SYK model as in other plots; $J=1$, $N=10$ (with $D=2^N = 1024$).}
\label{fig:SYKdosSFF}
\end{figure}

Being a Hamiltonian system, the SYK model satisfies $K(t) = \lvert \tildos(t)\rvert^2$, where $\dos(E)$ is the density of states as discussed in the main text [around Eq.~\eqref{eq:tildosdef}]. It has been noted~\cite{BlackHoleRandomMatrix, ShenkerThouless} that $\dos(E)$ has sharp edges for the SYK model, which leads to a power-law asymptotic behavior of the SFF. In particular, one obtains~\cite{BlackHoleRandomMatrix, ShenkerThouless} (see also Fig.~\ref{fig:SYKdosSFF})
\begin{equation}
K_{\text{SYK}}(t\to\infty) \sim \frac{\Theta(D^{-\alpha})}{t^3}+O(D^{-1}).
\label{eqs:SYKsff}
\end{equation}
Here, $\alpha$ is related to the zero-temperature entropy $\mathcal{S}_0$ of the SYK model~\cite{ShenkerThouless} via $e^{\mathcal{S}_0} = 2^{N(1-\alpha/2)}$, and behaves as $\alpha \propto q^{-2}$ for large $q$ with $\alpha \approx 0.66$ for $q=4$~\cite{MaldacenaStanford}. This power-law behavior is related to Statement 2 of the main text concerning a slower-than-exponential decay of the SFF in systems with finite spectral edges, with the slight difference that the spectral edges of the SYK model occur at asymptotically large energies in the $N\to\infty$ limit~\cite{ShenkerThouless}.

While numerical evidence for finite $N$ (Sec.~\ref{sec:SYKnumerics}, Fig.~\ref{fig:SYKscrambling}) suggests that subsystems of the SYK model scramble to $P_S(t) = D_S^{-1} + o(D_S^{-1})$ for the system size considered here, we will nevertheless derive the scrambling time limit for any $P_S(t) = O(D_S^{-1})$ to maintain generality (this can be extended to any $P_S(t) = O(D_S^{-\kappa})$ with similar qualitative conclusions). By the inequality Eq.~\eqref{eqs:dynamicalinequality}, this requires that $K(t > t_s) \leq O(D_S^{-1})$, which gives
\begin{align}
\frac{\Theta(D^{-\alpha})}{t_s^3} \leq O(D_S^{-1}) \implies t_s = \Omega\left[\left(\frac{D_S}{D^\alpha}\right)^{1/3}\right] = \Omega\left[2^{(N_S - \alpha N)/3}\right].
\label{eqs:SYKscramblingtime}
\end{align}
This yields Eq.~\eqref{eq:ts_SYK} of the main text when combined with the observation that we need $N_S > \alpha N$ for $t_s \to \infty$ (as we used the $t\to\infty$ asymptotics of the SFF), which is the regime in which this bound provides a nontrivial constraint; thus, the subsystems of interest must be at least a finite (but increasingly small with $q$) fraction of the system size, similar to the requirement in the definition of scrambling adopted in  Ref.~\cite{LashkariFastScrambling}.

As any basis for a subsystem (such as the $\proj_k$ in Eq.~\eqref{eqs:projdef}) must involve all energies by the completeness relation ($\sum_k \proj_k = \idop$), we are forced to include the entire spectrum including the edges and cannot eliminate the power-law decay by restricting to the center of the spectrum using, e.g., a Gaussian filter [$g(\hat{H}) \propto e^{-c\hat{H}^2}$], as done for this model in a different context in Ref.~\cite{ShenkerThouless}. While such filtering is mathematically possible in our inequality Eq.~\eqref{eqs:dynamicalinequality}, it is the physical choice of initial states that precludes this possibility (as noted in the discussion following Eq.~\eqref{eqs:finitetemp}) and leads to Eq.~\eqref{eqs:SYKscramblingtime}.

We should also emphasize that this power law decay entirely concerns the initial decay (``slope'') region of the SFF, and does not describe the behavior of the late-time quantum fluctuations (``ramp'' or ``plateau'', contained in the $O(D^{-1})$ term). The late-time regions are not particularly useful in setting rigorous speed limits on scrambling, for the following reasons. Firstly, $K(t) = O(D^{-1})$ in these regions, which is much less than $P_S(t) \sim D_S^{-1}$ for the infinite temperature scrambling of any subsystem of dimension $D_S \ll D$ (for which it is sufficient that $N_E = N-N_S \gg 1$, even if $N_S/N = \Theta(1)$). Secondly, a key observation in the study of many-body chaos (e.g., Ref.~\cite{ShenkerThouless}) is that thermalization generally occurs well before much of the ``ramp'' region in Hamiltonian systems including the SYK model; in fact, the time of onset of the quantum fluctuations corresponding to the ramp is comparable to the time required for complete thermalization in the full system in many cases~\cite{ShenkerThouless}, whereas smaller subsystems would thermalize much in advance of this time.

\subsection{Scrambling in the SYK model: Numerical details}
\label{sec:SYKnumerics}
Numerics were run using an iterative procedure (with increasing $N$) to generate a matrix representation for the $2N$ Majoranas $\hat{\chi}_j$ described in Ref.~\cite{SarosiSYKnumerics}, in a specific $D$-element orthonormal basis $\mathcal{B} = \lbrace\lvert B_k\rangle\rbrace_{k=1}^{D}$, from which the matrix elements of the Hamiltonian are readily obtained using Eq.~\eqref{eqs:defSYK}. 
Matrix representations of $\hat{c}_k$, $\hat{c}_k^\dagger$ are then obtained from Eqs~\eqref{eqs:cdef}, \eqref{eqs:cdaggerdef},
for which the vacuum state (here referring to the state with no fermion occupancy, annihilated by all the $\hat{c}_j$) is given by
\begin{equation}
\lvert 0\rangle = \lvert B_{1+\frac{D}{2}}\rangle.
\label{eqs:vaccumdef}
\end{equation}
This allows generating all the fermion Fock states in the $\mathcal{B}$-basis using Eq.~\eqref{eqs:Fockgendef}. The labels $1$ to $N$ of the fermion sites is fixed by the above procedure to generate the matrix elements of $\hat{c}_k$, $\hat{c}_k^\dagger$, for which we choose the first $N_S$ sites as our subsystem of interest as in Eq.~\eqref{eqs:subsysdef}. The initial basis states $\proj_k$ are given by Eq.~\eqref{eqs:projdef} re-expressed in the $\mathcal{B}$-basis.

We choose $q=4$, with $N=10$ fermion sites (or $2N = 20$ Majoranas) and $N_S = 7$ for numerical illustration ($D=1024$, $D_S = 128$, $D_E = 8$), with a randomly generated single realization of the disorder coefficients $J_{jk\ell m}$ in Eq.~\eqref{eqs:defSYK4}. The choice of $N_S = 7$ (so that $D_S = 1/128$) is because the first ``power-law'' oscillation of the SFF (after the initial decay) has a peak value of approximately $1/100$ for $N=10$, and we require $D_S^{-1} < 1/100$ to observe a nontrivial effect of this peak in delaying scrambling to $P_S(t) = D_S^{-1}$; we note that this choice also satisfies the condition $N_S > \alpha N$ in Eq.~\eqref{eq:ts_SYK} of the main text, with $\alpha \approx 0.66$ for $q=4$~\cite{MaldacenaStanford}. Higher values of $N$ or $N_S$ require significantly higher computational time due to exponentially growing complexity, whereas the above values are sufficient to provide a basic illustration of the bound of Eq.~\eqref{eqs:dynamicalinequality}.

\begin{figure}[!hbt]
\subfloat[][Linear-linear plot of $P_S(t)$, $K(t)$ and $\cos^2(\sigma_E t)$ (the latter for $0\leq \sigma_E t< \pi/2$), depicting the sensitivity required for many-body thermalization in large subsystems. The first $t^{-3}$ ``recurrence'' of the SFF around $t \approx 6$, whose nontrivial effect is clearly visible in the log-linear Fig. 1 of the main text, is barely discernible in this plot.]{\includegraphics[width=0.3 \columnwidth]{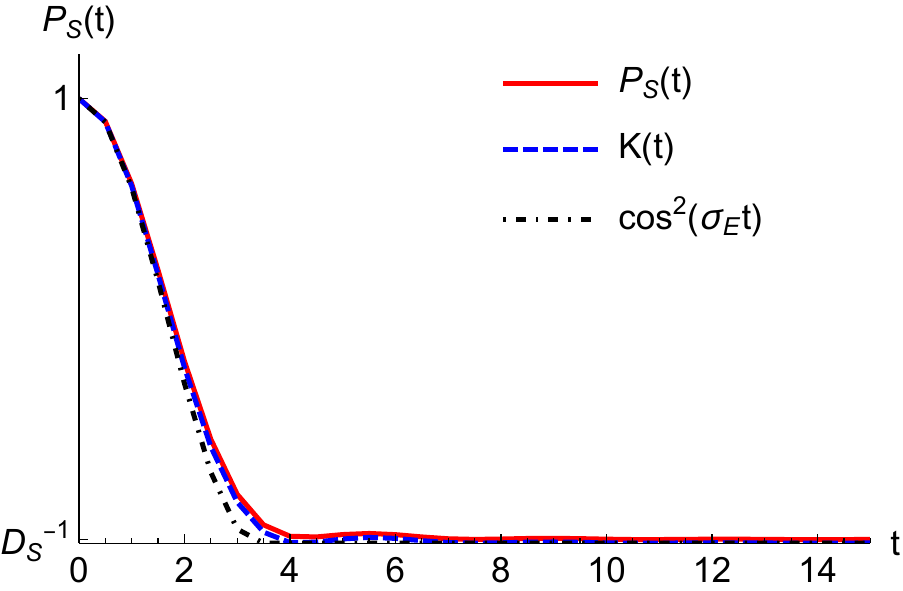} \label{fig:SYKlinplot}} \qquad
\subfloat[][Scaled mean return probability $D_S P_S(t)$ and return probabilities $D_S P_k(t)$ for the states defined in Eq.~\eqref{eqs:numericalstates}, suggesting that subsystems in this model satisfy $P_k(t\to \infty) \to 1/D_S$, and in particular $P_S(t\to\infty)\to 1/D_S$.]{\includegraphics[width=0.3 \columnwidth]{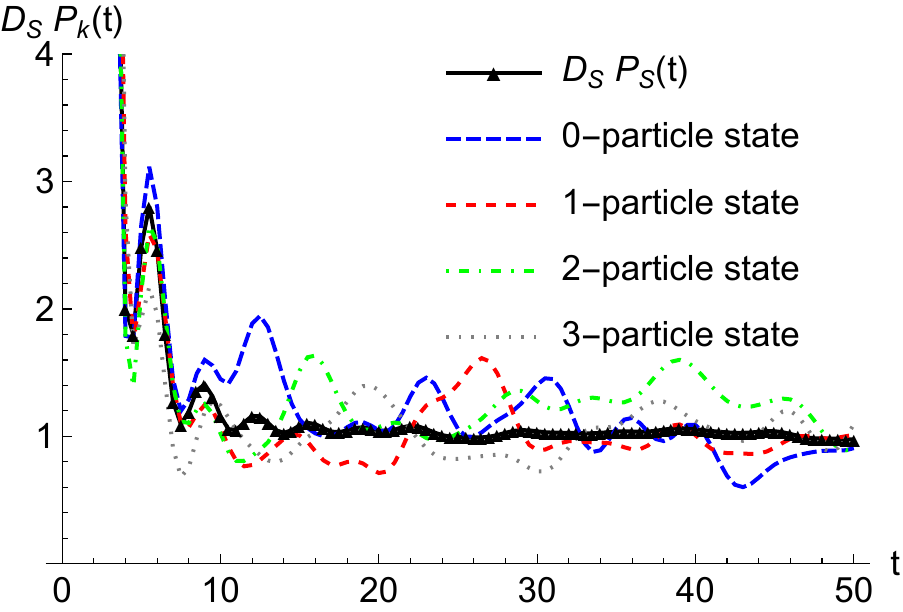} \label{fig:SYKretprobs}} \qquad
\subfloat[][Scaled overlaps $D_S Q_{kj}(t)$ between specific pairs of states defined in Eq.~\eqref{eqs:numericalstates}, suggesting that $Q_{kj}(t\to\infty) \to 1/D_S$, and therefore providing evidence that scrambling to infinite temperature in the sense of Eq.~\eqref{eqs:scramblingdef} does appear to occur in this model.]{\includegraphics[width=0.3\columnwidth]{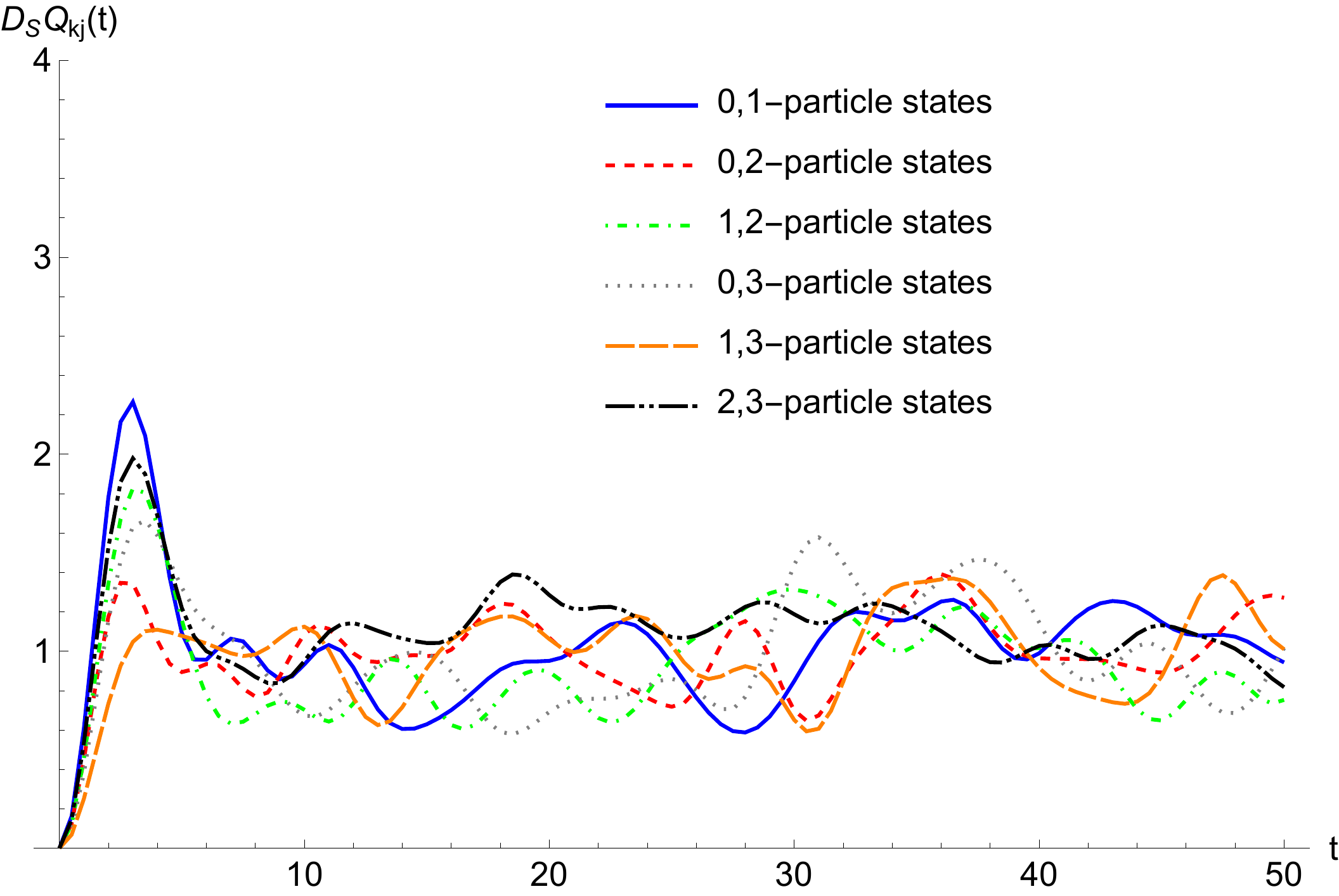} \label{fig:SYKoverlaps}} 
\caption{Additional numerics (a) demonstrating the high sensitivity of the speed limits such as $P_S(t) \geq K(t)$ required for a many-body system, and (b, c) providing evidence for scrambling to infinite temperature in fermion subsystems of the SYK model. All plots use the same realization of the interaction strengths $J_{jk\ell m}$ (each chosen from a Gaussian distribution as described near Eq.~\eqref{eqs:defSYK}) used in Fig. 1 of the main text.}
\label{fig:SYKscrambling}
\end{figure}

Fig. 1 of the main text illustrates our primary inequality Eq.~\eqref{eqs:dynamicalinequality}; here, we reproduce a linear-linear plot of the same data to highlight the sensitivity of the bound to microscopic fluctuations in the SFF and many-body thermalization [Fig.~\ref{fig:SYKlinplot}]. We also provide additional numerical evidence in Fig.~\ref{fig:SYKscrambling} of this supplement that the behavior of the SYK model appears to be consistent with the strong definition of scrambling adopted in the main text based on Ref.~\cite{LashkariFastScrambling} (corresponding to infinite temperature thermalization). For this, we consider the individual return probabilities,
\begin{equation}
P_k(t) \equiv \frac{1}{D_E}\Tr[\proj_k(t)\proj_k(0)],
\end{equation}
and overlaps between a time-evolved state and a distinct initial state:
\begin{equation}
Q_{kj}(t) \equiv \frac{1}{D_E}\Tr[\proj_k(t)\proj_j(0)].
\end{equation}
By Eq.~\eqref{eqs:scramblingdef} (or Eq.~\eqref{eq:scramblingdef} in the main text) both of these quantities should become $1/D_S$ to leading order. We see this to be the case in Figs.~\ref{fig:SYKretprobs}, \ref{fig:SYKoverlaps}. Further, the oscillations of $D_S P_k(t)$ and $D_S Q_{kj}(t)$ are consistent with $O(D_E^{-1/2})$ (in this case, $D_E^{-1/2} = 1/\sqrt{8} \approx 0.35$), as expected for the random-matrix-like thermalization behavior in Eq.~\eqref{eq:HaarScrambling} of the main text.

In Fig.~\ref{fig:SYKscrambling}, $P_S(t)$ refers to the mean return probability of all subsystem initial states of the form Eq.~\eqref{eqs:projdef}, while the specific other states involved in $P_k(t)$ and $Q_{kj}(t)$ are as follows:
\begin{align}
\text{0-particle state} &:\ \proj_{k0} = \lvert 0,0,0,0,0,0,0\rangle_S\langle 0,0,0,0,0,0,0\rvert \otimes \idop_E, \nonumber\\
\text{1-particle state} &:\ \proj_{k1} = \lvert 0,0,0,0,0,0,1\rangle_S\langle 0,0,0,0,0,0,1\rvert \otimes \idop_E, \nonumber\\
\text{2-particle state} &:\ \proj_{k2} = \lvert 0,0,0,0,0,1,1\rangle_S\langle 0,0,0,0,0,1,1\rvert \otimes \idop_E, \nonumber\\
\text{3-particle state} &:\ \proj_{k3} = \lvert 0,0,0,0,1,1,1\rangle_S\langle 0,0,0,0,1,1,1\rvert \otimes \idop_E.
\label{eqs:numericalstates}
\end{align}	
The terminology of ``$n$-particle states'' here only refers to the number of particles in the $N_S=7$ sites of the subsystem, while the number of particles in the remaining $N_E=3$ sites remains completely uncertain in these states (alternatively, each such projector corresponds to a set of all pure product states with a specific configuration in the subsystem, see Sec.~\ref{sec:pureproductstates}). We note that it is possible for these states to appear to completely ``mix'' into each other under time evolution and thereby appear to scramble to infinite temperature (as suggested by the numerics of Fig.~\ref{fig:SYKscrambling}), due to the lack of fermion number conservation in the SYK model.

\end{document}